\documentclass[12pt,a4paper]{article}
\usepackage[centertags]{amsmath}
\usepackage{amsmath,amsfonts,amssymb,amsthm,amstext,eucal,amsgen,amsbsy,amsopn}
\usepackage{newlfont}
\newlength{\defbaselineskip}
\newcommand{\al}{\alpha}
\newcommand{\be}{\beta}
\newcommand{\ga}{\gamma}
\newcommand{\de}{\delta}

\newcommand{\la}{\lambda}

\newcommand{\si}{\sigma}

\newcommand{\Ga}{\Gamma}
\newcommand{\La}{\Lambda}
\newcommand{\Om}{\Omega}

\newcommand{\eps}{\varepsilon}

\newcommand{\tht}{\tilde{\theta}}
\newcommand{\At}{\tilde{A}}

\newcommand{\at}{\tilde{a}}
\newcommand{\chit}{\tilde{\chi}}

\newcommand{\qh}{\hat{q}}
\newcommand{\bh}{\hat{b}}
\newcommand{\ch}{\hat{c}}
\newcommand{\ph}{\hat{p}}
\newcommand{\Lh}{\hat{L}}

\newcommand{\Qh}{\hat{Q}}

\newcommand{\Ah}{\hat{A}}
\newcommand{\Bh}{\hat{B}}
\newcommand{\Ch}{\hat{C}}

\newcommand{\Eh}{\hat{E}}
\newcommand{\Fh}{\hat{F}}
\newcommand{\Gh}{\hat{G}}
\newcommand{\Hh}{\hat{H}}
\newcommand{\Jh}{\hat{J}}
\newcommand{\Kh}{\hat{K}}
\newcommand{\Ph}{\hat{P}}
\newcommand{\Rh}{\hat{R}}
\newcommand{\Sh}{\hat{S}}
\newcommand{\Xh}{\hat{X}}

\newcommand{\Uh}{\hat{U}}
\newcommand{\Vh}{\hat{V}}
\newcommand{\lah}{\hat{\lambda}}
\newcommand{\wh}{\hat{w}}
\newcommand{\thetah}{\hat{\theta}}

\newcommand{\deeh}{\hat{d}}

\newcommand{\chih}{\hat{\chi}}

\newcommand{\etah}{\hat{\eta}}
\newcommand{\PPh}{\hat{\mathcal{P}}}
\newcommand{\thh}{\hat{\theta}}

\newcommand{\lab}{\bar{\lambda}}

\newcommand{\wb}{\bar{w}}
\newcommand{\psib}{\bar{\psi}}

\newcommand{\labh}{\hat{\bar{\lambda}}}
\newcommand{\wbh}{\hat{\bar{w}}}

\newcommand{\Imp}{\Rightarrow}

\newcommand{\preprox}{\approx}
\newcommand{\presim}{\simeq}

\newcommand{\Xd}{\dot{X}}

\newcommand{\ad}{\dot{a}}
\newcommand{\bd}{\dot{b}}

\newcommand{\PP}{\mathcal{P}}

\newcommand{\pa}{\partial}
\newcommand{\pab}{\bar{\partial}}

\newcommand{\half}{\frac{1}{2}}
\theoremstyle{plain}
\newtheorem{thm}{Theorem}[section]

\theoremstyle{definition}
\newtheorem{exa}[thm]{Example}
\theoremstyle{remark}

\numberwithin{equation}{section}

\begin{document}


\pagenumbering{arabic}

\title{Ghost constraints and the covariant quantization of the superparticle in ten dimensions}

\author{{\bf Michael Chesterman\footnote{\texttt{M.J.Chesterman@qmul.ac.uk}}}\\
{\small The Physics Department,}\\
{\small Queen Mary,}\\
\small{University of London,} \\
{\small Mile End Road, E1 4NS, UK}}

\date{December 2002\\
\small{Report no. QMUL-PH-02-23}}

\maketitle

\begin{abstract}
We present a modification of the Berkovits superparticle. This is
firstly in order to covariantly quantize the pure spinor ghosts,
and secondly to covariantly calculate matrix elements of a generic
operator between two states. We proceed by lifting the pure spinor
ghost constraints and regaining them through a BRST cohomology. We
are then able to perform a BRST quantization of the system in the
usual way, except for some interesting subtleties. Since the pure
spinor constraints are reducible, ghosts for ghosts terms are
needed, which have so far been calculated up to level 4. Even
without a completion of these terms, we are still able to
calculate arbitrary matrix elements of a physical operator between
two physical states.
\end{abstract}
\pagebreak

\tableofcontents

\section{Introduction}

The Brink-Schwarz superparticle action\cite{Brink:1981nb} yields a
manifestly super-Poincar\'e covariant, classical description of a
free particle moving in superspace. However, covariant
quantization has so far proved problematic.

Recently, Berkovits and collaborators have proposed a separate,
super-Poincar\'e covariant model for the $D=10$, $N=1$
superparticle \cite{Berkovits:2001rb,Berkovits:2002uc} which began
initially as a superstring model
\cite{Berkovits:2000fe,Berkovits:2000ph,
Berkovits:2000wm,Berkovits:2001mx,
Berkovits:2001ue,Berkovits:2001us, Berkovits:2002qx}. Also, in
\cite{Matone:2002ft}, the heterotic Berkovits string was derived
from the $n=2$ superembedding formulation, and an alternative
covariant approach without pure spinors was suggested in
\cite{Grassi:2001ug,Grassi:2002tz,Grassi:2002xv}, which we discuss
in more detail in section \ref{sec:gc3_alternative_approaches}.

The approach of Berkovits is derives from work by Howe
\cite{Howe:1991bx,Howe:1991mf}, in which an on-shell superspace
description of $D=10$ super Yang-Mills and supergravity is given
as integrability conditions along pure spinor lines. Berkovits
found a BRST operator, which we refer to as $\Qh$, with pure
spinors as ghosts. The ghost number one state cohomology of $\Qh$
is covariant and corresponds exactly with the quanta of on-shell
super-Maxwell theory, which is the correct spectrum for the
$D=10$, $N=1$ superparticle. While this is pleasing, there are
unsolved problems which we attempt to address in this article.

Firstly, there is a difficulty in finding a covariant description
of the physical degrees of freedom of the pure spinor ghost and
its conjugate momentum. Secondly, we require an inner product on
the Hilbert space. Thirdly, a systematic study of the space of
physical operators of the theory is needed, and a direct
comparison to the physical operators of the light-cone gauge
Brink-Schwarz superparticle should be made. Fourthly, there is an
issue that the Berkovits BRST operator is not hermitian. Our
findings on these problems are now discussed below.

\begin{enumerate}
\item Covariance: We argue later in section
\ref{sec:gc2_lambda_first_class} that the pure spinor constraints
should be treated as first class, i.e. as gauge generators. In
order to describe the physical degrees of freedom of the
constrained ghosts, one approach, as described in appendix
\ref{sec:sp7_U5}, is to completely gauge fix and solve the
combined pure spinor and gauge fixing constraints explicitly using
$U(5)$ co-ordinates, after first Wick-rotating from $SO(9,1)$ to
$SO(10)$. This approach has much in common with taking the
light-cone gauge for the bosonic particle. In both cases, only the
physical degrees of freedom remain after gauge fixing. Also,
Lorentz invariance is broken, in the case of pure spinors from
$SO(10)$ to $U(5)$, and in the case of the particle to $SO(8)$.

It seems natural, just as with the bosonic particle, to attempt a
BRST approach in which instead of removing unphysical degrees of
freedom, the phase space is expanded with extra ghosts thus
maintaining covariance. Physical operators and states are then
regained through a BRST cohomology.

There are however extra subtleties involved in imposing ghost
constraints as opposed to ordinary constraints. In particular, the
ghost constraints cannot be combined into the Berkovits BRST
operator. We find that the solution is to introduce a second BRST
operator, $\Qh_{gc}$, which simply implements the pure spinor
constraints. The Berkovits BRST operator, $\Qh$, then becomes
nilpotent modulo $\Qh_{gc}$. The pair of BRST operators form what
is known as a BRST double complex.

The pure spinor constraints are reducible, thus ghost for ghost
terms are required. So far these terms have been calculated up to
level 4. As the number of ghosts for ghosts increases level by
level, it seems likely that infinitely many terms will be
required. This is not definite though as no pattern has been
found, and there is no unique choice of term at each level.
Despite this problem, we find that by choosing a suitable
representative from each cohomology class, the ghost part factors
out in any matrix element calculation of a physical operator
between two states. Also, the state cohomology is not affected by
this difficulty.

\item The inner product: Using the natural Schr\"odinger measure,
the norm of the Berkovits wavefunction is zero, so the question
arises of how to define the inner product. Actually, this
situation is normal for BRST quantization with no minimal sector
using the Schr\"odinger representation. Take the bosonic particle
for example. States in the Berkovits cohomology couple in the
inner product to states in a dual cohomology at opposite ghost
number. The solution is thus to find a map between the two
cohomologies and place states of one cohomology on the left and
states of the other on the right within the inner product. In the
case of the bosonic particle, in order to obtain the dual ghost
number $1/2$ wavefunction, you multiply the corresponding ghost
number $-1/2$ wavefunction by the $c$ ghost. In general, however,
there is no simple map like this, and the dual cohomology has to
be explicitly calculated.

\item The operator cohomology: The Berkovits operator cohomology
does not correspond with the physical operators of the
Brink-Schwarz superparticle. However, we discover that $\Qh$
indirectly implies `effective constraints', which are not
obviously present in the Berkovits BRST operator. These are simply
related to the on-shell equations for super Yang-Mills. The
operator cohomology modulo these effective constraints matches the
physical operators of the Brink-Schwarz superparticle. Thus, the
Berkovits and Brink-Schwarz superparticles are equivalent. We also
find that these effective constraints are the first class
constraints of the Brink-Schwarz particle.

\item Non-hermicity of $\Qh$: Naturally, this can only become an
issue once we have defined an inner product. The solution is found
in the definition of the inner product, or rather of the dual
cohomology.
\end{enumerate}

It should be noted that while issues 2), 3) and 4) are all solved
using 1), in principle they can also be studied using $U(5)$
co-ordinates as in appendix \ref{sec:sp7_U5}.

The paper is structured as follows. In section
\ref{sec:sp6_berk_superp}, we briefly review the Berkovits
superparticle model, in section \ref{sec:gc2_lambda_first_class}
we introduce the idea of the purity constraints being first class,
in section \ref{sec:gc5_formal} the general BRST formulation for
theories with first class ghost constraints is detailed. In
section \ref{sec:gc6_examples}, a simple example with linear ghost
constraints is given, in section \ref{sec:gc3_pre_BRST_operator},
we show how $\Qh_{gc}$ is constructed to 4th level. In section
\ref{sec:gc4_superp}, we finally build the superparticle model. We
also make an analogy with Chern-Simons theory, and compare with
the light-cone gauge Brink-Schwarz superparticle. In section
\ref{sec:gc7_quant_open_string}, we show how our covariant method
leads to anomaly cancellation for the open superstring and in
section \ref{sec:gc8_future_research} we discuss plans for future
research. The appendices mostly consist of relevant reference
material. However, note that appendix \ref{sec:sp7_U5} on the
description of pure spinors using U(5) co-ordinates, is different
to the usual Berkovits approach.

\section{The D=10, N=1 Berkovits Superparticle}\label{sec:sp6_berk_superp}

The Berkovits, BRST invariant superparticle action
\cite{Berkovits:2002uc}, which is in Hamiltonian form, is given by
\begin{equation}\label{eq:sp6_action}
S_B=\int{d\tau\, (\Xd^m P_m+ \dot{\theta}^\al p_\al  +
\dot{\la}^\al w_\al - \frac{1}{2}P_m P^m)},
\end{equation}
where variables $X^m$, $\theta^\al$ are the usual $D=10$, $N=1$
superspace co-ordinates, and $P_m$, $p_\al$ their conjugate
momenta with $m=1\ldots 10$ and $\al=1\ldots 16$. Also,
$\theta^\al$ and $p_ \al$ are fermionic, Majorana-Weyl spinors of
opposite chirality. The ghosts $\la^\al$ and $w_\al$ are
bosonic, complex, Weyl spinors with ghost numbers $1$ and $-1$
respectively. The notation used for $D=10$ spinors and gamma
matrices is described in appendix \ref{sec:gamma_matrices}.

Berkovits defines a BRST operator
\begin{eqnarray}\label{eq:sp6_Q_sp}
\Qh = \lah^\al \deeh_\al, && \Qh^2 = \Ph_m \lah^\al \ga^m_{\al\be}
\lah^\be,
\end{eqnarray}
where $\deeh_\al= \ph_\al - i\Ph_m (\ga^m\thetah)_\al$ are the
fermionic constraint functions of the Brink-Schwarz superparticle,
and where $\lah^\al$ are defined to obey `purity' constraints
\begin{eqnarray}\label{eq:sp6_pure}
\lah^\al \ga^m_{\al\be} \lah^\be = 0,
\end{eqnarray}
in order that $\Qh$ be nilpotent. As shown in appendix
\ref{sec:sp7_U5}, the purity constraints \eqref{eq:sp6_pure} leave
$\la^\al$ with 11, complex degrees of freedom.

Given ghost number operator $\Gh'=i\lah^\al \wh_\al$, the ghost
number one, state cohomology $H^1_{st}(\Qh)$ describes the
physical modes of super-Maxwell theory in a superspace covariant
way. Using the Schr\"odinger representation, we find
\begin{eqnarray}\label{eq:sp6_BRST}
Q\psi = 0, && \de\psi = Q \phi(X,\theta)\\
\Imp \ga_{mnpqr}^{\al\be} D_\al A_\be =0, && \de A_\al = D_\al
\phi \label{eq:gc2_S_Max}
\end{eqnarray}
where $\psi = \la^\al A_\al(X,\theta)$ and $\phi(X,\theta)$ are
generic ghost number one, and ghost number zero wavefunctions
respectively, and where $D_\al$ is the usual covariant superspace
derivative
\begin{eqnarray}\label{eq:gc2_cov_derivative}
D_\al \equiv \frac{\pa}{\pa \theta^\al} - i\ga^m_{\al\be}
\theta^\be \pa_m, && \deeh_\al \equiv -iD_\al.
\end{eqnarray}
We have also used the identity
\begin{equation}
\la^\al \la^\be = \frac{1}{16}\la\ga^m \la \ga_m^{\al\be} +
\frac{1}{16\cdot 5!}\la \ga^{mnpqr}\la \ga_{mnpqr}^{\al\be},
\end{equation}
which comes from equation \eqref{eq:C_bispinor}, recalling that
$\ga_{mnp}^{\al\be}$ is antisymmetric in $\al$ and $\be$, and so
does not contribute. Equation \eqref{eq:gc2_S_Max} describes the
equations of motion and gauge transformation for $D=10$, $N=1$
super-Maxwell theory. Therefore $H^1_{st}(\Qh)$ corresponds
exactly with the spectrum of the Brink-Schwarz superparticle in
the light-cone gauge.

\section{Pure spinor ghost constraints as first class Dirac
constraints}\label{sec:gc2_lambda_first_class}

\subsection{The BRST double complex}

In order to obtain the equations of motion for Berkovits'
superparticle, we should solve $\de S_B=0$ on the constraint
surface $\la\ga^m\la$=0. Equivalently, we define an action
\begin{equation}\label{eq:gc4_S_intermediate}
S=\int{d\tau\,(\dot{X}^m P_m  + \dot{\theta}^\al  p_\al +
\dot{\la}^\al w_\al - \frac{1}{2}P_m P^m - \La_m \la\ga^m\la)},
\end{equation}
where $\La_m$ are ghost number $-2$, Lagrange multipliers, and
then solve $\de S=0$ globally. Since $[\la\ga^m\la,
\la\ga^n\la]=0$, $S$ has the gauge symmetries
\begin{eqnarray}
\de_\eps \la^\al = 0, & \de_\eps w_\al = -2 \eps_m (\ga^m\la)_\al,
& \de_\eps \La_m = \dot{\eps}_m,
\end{eqnarray}
where $\eps_m(\tau)$ is a local, bosonic, ghost number $-2$
parameter. Thus, the purity constraints can be interpreted as first
class constraints. Observables should be gauge invariant with
respect to the ghost constraints, as well as $\Qh$-closed, as
already argued by Berkovits \cite{Berkovits:2002uc}.

In order to covariantly quantize the ghosts, a BRST implementation
of the gauge generators $\la\ga^m\la$ is required. We define a
separate BRST operator $\Qh_{gc}$, with its own associated
anti-hermitian ghost number operator $\Gh_{gc}$
\begin{eqnarray}
\Qh_{gc} = \Ch_m \lah\ga^m\lah + \ldots, && \Gh_{gc}=\frac{i}{2}
(\Ch_m\Bh^m - \Bh^m \Ch_m +\ldots),
\end{eqnarray}
where $\Ch_m$ and $\Bh^m$ are a fermionic conjugate pair of
ghosts. Also, the ellipses refer to ghost for ghost terms, which
are needed because the pure spinor constraints are reducible.
Calculation of these terms is discussed in section
\ref{sec:gc3_pre_BRST_operator}. It should be noticed that the
Berkovits ghosts $\la^\al$ and $w_\al$ commute with $\Gh_{gc}$,
and hence from from the point of view of $\Qh_{gc}$ are treated as
ordinary ghost number zero variables. Using this approach,
$\la^\al$ and $w_\al$ are unconstrained and the pure spinor
constraints are realized through requiring physical operators and
states to belong to the operator or state cohomology
$H(\Qh_{gc})$.

Since $\la^\al$ are unconstrained, $\Qh$ is no longer nilpotent.
However, we say that $\Qh$ is a BRST operator modulo $\Qh_{gc}$,
since
\begin{eqnarray}\label{eq:gc3_Q_BRST}
[\Qh,\Qh_{gc}] = 0, && {\Qh}^2 = [-i\Bh^m \Ph_m , \Qh_{gc}].
\end{eqnarray}
The first equation of \eqref{eq:gc3_Q_BRST} implies that $\Qh$
maps any cohomology class of $H(\Qh_{gc})$ onto another one, and
the second that $\Qh$ is nilpotent within the phase-space defined
by the cohomology $H(\Qh_{gc})$. Physical operators and states are
given by the cohomology of $\Qh$ within the cohomology of
$\Qh_{gc}$, which is denoted as $H(\Qh |H(\Qh_{gc}))$.

A pair of operators obeying the same general algebra as $\Qh$ and
$\Qh_{gc}$ is known as a double complex. This construction is
common in mathematics, for example in the calculation of
equivariant integrals. However, except for a mention in
\cite{Batalin:1992ag}, this type of double complex does not seem
to have been explored before in the context of two BRST operators.

In the remainder of this article, we describe how to implement
this BRST double complex for a generic constrained ghost system,
and in particular for the Berkovits superparticle.

\subsection{Discussion of alternative approaches}\label{sec:gc3_alternative_approaches}

As a non-covariant alternative to quantizing the pure spinor
ghosts, we show in appendix \ref{sec:sp7_U5}\footnote{Our approach
to U(5) co-ordinates differs from that of Berkovits
\cite{Berkovits:2000fe,Berkovits:2001us}, which isn't equivalent
to a canonical gauge fixing of the first class pure spinor
constraints, at least not with initial Poisson bracket
$[\la^\al,w_\be]=\de^\al_\be$. Our approach has the advantage of
not requiring the fermionization of ghosts.} how to construct
canonical, gauge-fixing constraints\cite{Henneaux:1992ig}, using
$U(5)$ co-ordinates, which completely fix the gauge symmetry
generated by $\la\ga^m\la$ in a certain region $\la^+ \neq 0$,
where $\la^+$ is defined in the appendix. This is similar in
approach to imposing the light-cone gauge for the bosonic particle
for example. There is an obvious disadvantage here that the
Lorentz covariance is reduced to $U(5)$ covariance and also that
it's not valid for $\la^+=0$.

One might expect to be able to build a single covariant BRST
operator $Q$ with single ghost number, which implements both the
Berkovits BRST operator \eqref{eq:sp6_Q_sp} and the pure spinor
constraints
\begin{equation}\label{eq:gc4_wrongQ}
Q = \la^\al d_\al + b_m\la\ga^m\la + \ldots,
\end{equation}
where $b_m$ are fermionic ghosts with ghost number -1 and where
the dots refer to `closure' terms, whose sole purpose is to ensure
nilpotence of $Q$ off-shell, as opposed to `physical' terms, which
specify the gauge generators. However, it is impossible for $Q$ to
be nilpotent without $b_m\la\ga^m\la$ becoming a `closure' term
and without introducing new, `physical', gauge generator terms,
which change the physical nature of the theory. It is then wrong
to think of this term as introducing ghost constraints, since
constraints are implemented only in the `physical' terms.

Having said this, in the approach taken by Van Nieuwenhuizen and
collaborators \cite{Grassi:2001ug,Grassi:2002tz,Grassi:2002xv} for
the covariant quantization of the superstring, they essentially
begin with the above BRST operator \eqref{eq:gc4_wrongQ} and
through some procedure introduce new `physical' and `closure'
terms until $Q$ becomes nilpotent. Now $b_m\la\ga^m\la$ becomes a
`closure' term, thus there are no longer pure spinor constraints.
The resultant BRST operator isn't directly equivalent to the
Berkovits BRST operator due to the extra gauge generators, in fact
its cohomology is null. Remarkably though, by restricting physical
Vertex operators further to a certain subspace of all possible
operators, the cohomology on this subspace, known as an
equivariant cohomology, has been shown to be equivalent to the
Berkovits cohomology for the open superstring, at least in the
massless sector. The advantage of this approach is that there are
no pure spinors, and hence this bypasses the problem of a
covariant description of them. There is however an issue which
needs clarification, which is why there is no central charge
cancellation in ten dimensions.

\section{Formal description of the method for arbitrary, first
class ghost constraints}\label{sec:gc5_formal} We now describe the
general formulation for the operator quantization of a BRST system
with first class ghost constraints. A path integral formulation is
also provided in \cite{Chesterman:2002a}. No attempt will be made
at this stage to interpret the constrained ghost quantum system,
nor to determine suitable ghost constraints in the general case,
since this depends on the particular system in question. Rather,
we assume that the defining operators, which are the BRST
invariant, bosonic Hamiltonian $\Hh$ and the two, fermionic BRST
operators $\Qh$ and $\Qh_{gc}$, have already been constructed and
we proceed to build the generic quantum system from them. Given
ghost constraint, BRST operator $\Qh_{gc}$ and corresponding ghost
number operator $\Gh_{gc}$
\begin{eqnarray}\label{eq:gc5_Q_pre_squared}
\Qh_{gc}^2 =0, && [\Gh_{gc},\Qh_{gc}]=\Qh_{gc},
\end{eqnarray}
the operators $\Qh$ and $\Gh$ are $\Qh_{gc}$-closed, since they
must map between cohomology classes of $H(\Qh_{gc})$
\begin{eqnarray}\label{eq:gc5_Q_ph_G_ph_Q_pre_closed}
[\Qh,\Qh_{gc}] = 0, && [\Gh,\Qh_{gc}] = 0.
\end{eqnarray}
Also
\begin{eqnarray}\label{eq:gc4_Q_nilpotent}
\Qh^2 \presim 0, && [\Gh, \Qh] \presim \Qh,
\end{eqnarray}
where
\begin{eqnarray}
\Ah \presim \Bh &\Imp& \Ah = \Bh + [\Ch,\Qh_{gc}],
\end{eqnarray}
for some operator $\Ch$. We will only consider theories where
$\Gh$ and $\Gh_{gc}$ commute, so that states can have both ghost
numbers well-defined,
\begin{equation}\label{eq:gc5_ghost_op_commute}
[\Gh,\Gh_{gc}] = 0.
\end{equation}

Physical operators belong to the ghost number zero, operator
cohomology of $\Qh$ within the ghost constraint ghost number zero,
operator cohomology of $\Qh_{gc}$, which we denote
$H^0_{op}(\Qh|H^0_{op}(\Qh_{gc}))$. A physical operator $\Vh$ thus
obeys
\begin{eqnarray}\label{eq:gc4_V_physical1}
[\Vh, \Qh_{gc}]=0, && [\Vh, \Qh] \presim 0\\
{[}\Vh, \Gh_{gc}]=0, && [\Vh, \Gh] \presim 0.
\label{eq:gc4_V_physical2}
\end{eqnarray}
There are two types of BRST-exact term that one can add to $\Vh$
in a BRST transformation
\begin{eqnarray}\label{eq:gc5_Qph_exact}
\Vh \sim \Vh + [\Uh,\Qh] +[\Uh_{gc},\Qh_{gc}] && \text{given }
[\Uh,\Qh_{gc}]=0.
\end{eqnarray}

Similarly, a physical state $\psi$ belongs to the ghost number
$g$, state cohomology of $\Qh$ within the ghost constraint ghost
number $k$, state cohomology of $\Qh_{gc}$, which we denote
$H^g_{st}(\Qh|H^{k}_{st}(\Qh_{gc}))$. Thus, $\psi$ satisfies
\begin{eqnarray}\label{eq:gc5_phys_state}
\Qh_{gc}\psi = 0, && \Qh\psi \presim 0\\
\Gh_{gc}\psi = k\psi, && \Gh \psi \presim g\psi,
\label{eq:gc5_phys_state2}
\end{eqnarray}
where the ghost numbers $g$ and $k$ depend on various factors, for
instance whether a Schr\"odinger or Fock representation is being
used. Again, there are two types of BRST-exact terms that one can
add to $\psi$ in a BRST transformation
\begin{eqnarray}
\psi \sim \psi + \Qh\chi + \Qh_{gc}\chi_{gc} && \text{given }
\Qh_{gc}\chi=0.
\end{eqnarray}

The BRST invariant Hamiltonian $\Hh$ is a physical, hermitian
operator which belongs to $H^0_{op}(\Qh|H^0_{op}(\Qh_{gc}))$, and
thus obeys equations \eqref{eq:gc4_V_physical1} and
\eqref{eq:gc4_V_physical2}. It is uniquely defined up to
BRST-exact terms
\begin{eqnarray}\label{eq:gc5_H_exact}
\Hh \sim \Hh + [\chih_{gc} , \Qh_{gc}] + [\chih, \Qh], &&
\text{where } [\chih,\Qh_{gc}] = 0,
\end{eqnarray}
where $\chih_{gc}$ and $\chih$ are gauge-fixing fermions.

In general we require
\begin{eqnarray}\label{eq:gc4_Q_hermitian}
\Qh_{gc}=\Qh_{gc}^\dag && \Qh=\Qh^\dag,
\end{eqnarray}
in order that if $\Qh_{gc}\psi=0$, then $(\psi, \Qh_{gc}\phi)=0$
for arbitrary state $\phi$, where $(\cdot,\cdot)$ is the inner
product on the Hilbert space, and where a similar result applies
for $\Qh$. However, in the special case of the Berkovits
superparticle $\Qh \neq \Qh^\dag$ and $\Qh_{gc} \neq
\Qh_{gc}^\dag$, since $\lah^\al \neq \lah^{\al\dag}$ because
$\la^\al$ are classically complex. This problem is solved in
section \ref{sec:gc7_phys_state_coho}, in short by taking the
complex conjugate of a wavefunction before placing it to the left
in the inner product.

The most important consequence of \eqref{eq:gc4_Q_hermitian} is
that BRST exact operators have vanishing matrix elements between
physical states
\begin{eqnarray}\label{eq:gc4_Q_pre_exact_ampl}
(\psi_m,[\Uh_{gc},\Qh_{gc}] \psi_n) = 0, && \text{for } \psi_m
\in H^{k}_{st}(\Qh_{gc})\\
(\psi_m,[\Uh,\Qh]\psi_n) = 0, &&\text{for } \psi_m \in
H^g_{st}(\Qh|H^{k}_{st}(\Qh_{gc})),\label{eq:gc4_Q_ph_exact_ampl}
\end{eqnarray}
where we have also used that $[\Uh,\Qh_{gc}]=0$ and equation
\eqref{eq:gc5_phys_state}.

As a final observation, any operator $\Ah$ belonging to
$H_{op}(\Qh_{gc})$ can be meaningfully expressed as the matrix
$(\psi_m, \Ah \psi_n)$, where $\{\psi_m\}$ form a basis for
$H_{st}(\Qh_{gc})$. When equations involving such operators are
written in matrix form, then the symbol $\presim$ can be replaced
with an equals sign. For example,
\begin{eqnarray}
(\psi_m, {\Qh}^2 \psi_n) = 0, && \text{given }\psi_m \in
H_{st}(\Qh_{gc}).
\end{eqnarray}


\section{An example of linear ghost constraints}\label{sec:gc6_examples}

We now illustrate the formal description of the last section with
a simple example. We show how a BRST system with linear ghost
constraints, specified by $\Qh_{gc}$, can be related to a gauge
theory with a single BRST operator. \exa[A simple example]{
Consider the motion of a particle described by the action
\begin{equation} S=\int_{t_1}^{t_2}{dt\, (-\half
(q^1)^2 + \half (\dot{q}^2)^2 - \half (q^2)^2)}.
\end{equation}
The Dirac-Bergmann algorithm yields second class constraints
\begin{eqnarray}
p_1=0, && q^1=0,
\end{eqnarray}
where the first is primary and the second is secondary.
The canonical Hamiltonian is
\begin{equation}
H=\half (q^1)^2 + \half (p_2)^2 + \half (q^2)^2.
\end{equation}
By using the Dirac bracket, or simply by parameterizing the
constraint surface using only co-ordinates $(q^2,p_2)$,
quantization is straightforward.

However, if we were to naively treat the constraints as if
they were first class, we could construct the BRST operator
\begin{eqnarray}
\Qh=\etah^1 \qh^1 + \etah^2 \ph_1, &&  \Qh^2 = i \etah^1\etah^2,
\end{eqnarray}
where $(\etah^1,\PPh_1)$ and $(\etah^2,\PPh_2)$ are fermionic
ghost, ghost momenta pairs. Supposing the first class ghost
constraint $\etah^1 =0$ is introduced for example, by defining
nilpotent BRST operator $\Qh_{gc}$, and its corresponding
anti-hermitian ghost operator $\Gh_{gc}$
\begin{eqnarray}
\Qh_{gc}=\Uh \etah^1, && \Gh_{gc}=\frac{i}{2}(\Uh \Vh + \Vh \Uh)
\end{eqnarray}
where $\Uh$ and $\Vh$ define a bosonic, ghost conjugate pair. The
anti-hermitian, ghost number operator $i/2(\etah^1\PPh_1 -
\PPh_1\etah^1 + \etah^2\PPh_2 - \PPh_2\etah^2)$ is gauge invariant
with respect to constraint $\etah^1=0$, but requires a BRST
extension in order to become $\Qh_{gc}$-closed
\begin{equation}
\Gh = \frac{i}{2}(\etah^1\PPh_1 - \PPh_1\etah^1 + \etah^2\PPh_2 -
\PPh_2\etah^2 - (\Uh \Vh + \Vh \Uh)).
\end{equation}
We can now verify that the above definitions of $\Qh_{gc}$, $\Qh$,
$\Gh_{gc}$ and $\Gh$ obey the required equations
\eqref{eq:gc5_Q_pre_squared} to \eqref{eq:gc5_ghost_op_commute}.

Now let us calculate the classical, physical functions. Since $Q
\presim \eta^2 p_1$, we deduce that
\begin{equation}\label{eq:gc5_dbrst_to_sbrst}
H^0(Q|H^0(Q_{gc})) \cong H^0(\eta^2 p_1 |
C^{\infty}(q^i,p_i,\eta^2,\PP_2)) \cong C^{\infty}(q^2,p_2).
\end{equation}
In effect, the ghosts $U$ and $V$ have cancelled the ghosts
$\eta^1$ and $\PP_1$, thus establishing equivalence to an ordinary
BRST theory, with BRST operator $\eta^2 p_1$ in phase-space
defined by canonical co-ordinates $(q^i,p_i,\eta^2,\PP_2)$.

In this particular example, introducing the ghost constraint
$\eta^1=0$ is equivalent to removing the second class constraint
$q^1=0$, which one can think of as a canonical gauge-fixing
constraint, thus leaving only first class constraint $p_1=0$. The
process of removing canonical gauge-fixing constraints is
sometimes known as `gauge unfixing'
\cite{Harada:1990aj,Mitra:1990mp,Vytheeswaran:1999da}, which is
also similar to `split involution' \cite{Batalin:1992ag}. There is
some similarity between our approach here and the recent
projection operator approach to the BRST quantization of general
constrained systems by Batalin et al. \cite{Batalin:2001dk}. There
they also introduce ghosts for all the second class constraints,
for the purpose of covariance. The extra ghost degrees of freedom
are then cancelled by adding ghost for ghost terms to the BRST
operator.

The Hamiltonian isn't BRST invariant, since it isn't gauge
invariant with respect to $p_1$ because $[H,p_1]=-q^1$ isn't zero
on $p_1=0$. Thus, we replace $H$ with gauge invariant Hamiltonian
$\tilde{H}$, which possesses the properties
\begin{eqnarray}
\tilde{H}|_{q^1=p_1=0} = H|_{q^1=p_1=0}, &&
[\tilde{H},p_1]|_{p_1=0} = 0.
\end{eqnarray}
A suitable choice is
\begin{equation}
\tilde{H}=\half (q^2)^2 + \half (p_2)^2,
\end{equation}
which is $Q_{gc}$-closed without need for further BRST extension.

We now calculate states and operators in the Schr\"odinger
representation in order to observe how the classical equivalence
shown in equation \eqref{eq:gc5_dbrst_to_sbrst}, is extended to a
quantum mechanical equivalence. It is simpler to notice first that
\begin{eqnarray}
\Gh  \presim \frac{i}{2}(\etah^2\PPh_2 - \PPh_2\etah^2), && \Qh
\presim \etah^2\ph_1.
\end{eqnarray}
The `physical' states in the ghost constraint cohomology appear at
ghost constraint ghost numbers -1/2 and +1/2 defined by states
$\psi_{U=0}$ and $\psi_{V=0}$ respectively, where in this
notation,
\begin{eqnarray}
\Uh \psi_{U=0} = 0, & \Vh \psi_{V=0}=0, & (\psi_{U=0}, \psi_{V=0})
= 1.
\end{eqnarray}
Thus, a basis for $\Qh_{gc}$-closed wavefunctions with ghost
constraint ghost numbers $\pm 1/2$ is given by
\begin{eqnarray}
\psi_{-1/2}(U,\eta^1) = \de(U)(a\eta^1+b), &&
\psi_{1/2}(U,\eta^1)=c\eta^1,
\end{eqnarray}
where $a$, $b$ and $c$ are c-number constants. Note that $a$ can
be transformed to zero by adding $\Qh_{gc}$-exact state $\Qh_{gc}
\de'(U)a$ to $\psi_{1/2}$. A generic, physical wavefunction $\psi
\in H^{\pm\half}_{st}(\Qh|H^{\pm \half}(\Qh_{gc}))$ can be written
as
\begin{eqnarray}
\psi= \psi_{\pm \half}(U,\eta^1) \psi_m(\eta^2,q^1,q^2),
&\text{given}& \Qh \psi_m(\eta^2,q^1,q^2)=0.
\end{eqnarray}

All operators $\Fh \in H^0_{op}(\Qh|H^0_{op}(\Qh_{gc}))$ can be
written in the form
\begin{equation}
\Fh=\Fh_1(\etah^2,\PPh_2,\qh^i,\ph_i)+
\Fh_2(\Uh,\Vh,\etah^i,\PPh_i,\qh^i,\ph_i) \presim
\Fh_1(\etah^2,\PPh_2,\qh^i,\ph_i).
\end{equation}
This is because $\Fh_1$ is separately $\Qh_{gc}$-closed, being
independent of $\Vh$ and $\PPh_1$, and since $H^0(Q_{gc}) \cong
C^{\infty}(q^i,p_i,\eta^2,\PP_2)$, $\Fh_2$ must be
$\Qh_{gc}$-exact. Therefore the most general matrix element of
physical operator $\Fh$ between two physical states is
\begin{equation}
\begin{split}
(\psi_{\half}(U,\eta^1)\psi_m(\eta^2,q^1,q^2), \,\Fh
\psi_n(\eta^2,q^1,q^2)\psi_{-\half}(U,\eta^1))\\
=(\psi_m(\eta^2,q^1,q^2), \,
\Fh_1(\etah^2,\PPh_2,\qh^i,\ph_i)\psi_n(\eta^2,q^1,q^2)),
\end{split}
\end{equation}
given $b=c=1$, where to obtain the second line, we have integrated
out $\eta^1$ and $U$ in the Schr\"odinger inner product. Thus, we
have seen how quantum mechanically, the above system is equivalent
to a gauge theory with single BRST operator $\Qh=\etah^2\ph_1$ and
phase space co-ordinates $(\qh^i,\ph_i,\etah^2,\PPh_2)$. }
\section{The pure spinor BRST operator $\Qh_{gc}$}\label{sec:gc3_pre_BRST_operator}

As already mentioned in section \ref{sec:gc2_lambda_first_class},
the pure spinor constraints $\la\ga^m\la=0$ are reducible and
hence $\Qh_{gc}$ requires ghost for ghost terms. Unfortunately,
only terms up to the fourth level of ghosts for ghosts have been
found so far. It is not yet known whether a covariant termination
will exist or whether infinite ghosts for ghosts will be needed.
In section \ref{sec:gc4_superp}, we nominally specify $\Qh_{gc}$
up to the 2nd level of ghosts, which is just high enough in order
to spot any patterns in the BRST extension of $\Qh_{gc}$-closed
operators. Without a full solution, there is no advantage in
specifying $\Qh_{gc}$ to the highest known level of ghosts.

We proceed with a brief recipe describing how to build a generic
BRST operator with reducible constraints \cite{Henneaux:1992ig},
and then build the reducibility identities up to level 4 and
$\Qh_{gc}$ up to level 2.

\subsection{A recipe for the construction of a BRST charge with
reducible constraints}\label{sec:cs3_reducible_BRST}

We begin with a set of first class constraints
\begin{eqnarray}
g_{a_0} = 0 && a_0 = 1, \ldots, m_0,
\end{eqnarray}
of which $m \leq m_0$ are independent. We define $Z_1$ with the
properties
\begin{eqnarray}\label{eq:gc6_Zg}
{(Z_1)_{a_1}}^{a_0}g_{a_0} =& 0, & a_1 = 1, \ldots, m_1\\
\text{Rank } {(Z_1)_{a_1}}^{a_0} \approx& m_0-m,
\end{eqnarray}
where $A \approx B$ implies that $A$ is equal to $B$ on the
constraint surface $g_{a_0}=0$. So $Z_1$ not only annihilates
$g_{a_0}$ globally in phase-space, but also describes all
$(m_0-m)$ vanishing linear combinations of the gauge generators
locally on the constraint surface, since ${(Z_1)_{a_1}}^{a_0}
[g_{a_0}, F]\approx 0$, for any function $F$. The order $k$
reducibility identity describes all vanishing linear combinations
of $Z_{k-1}$
\begin{eqnarray}\label{eq:gc6_ZZ_0}
{(Z_k)_{a_k}}^{a_{k-1}} {(Z_{k-1})_{a_{k-1}}}^{a_{k-2}} \approx 0,
&& a_k=1,\ldots m_k.
\end{eqnarray}
where
\begin{eqnarray}\label{eq:gc6_rank}
\text{Rank } {(Z_k)_{a_k}}^{a_{k-1}} \approx m_k', && m_k' =
m_{k-1} - m_{k-1}'.
\end{eqnarray}
We keep building more $Z_k$'s until there are no vanishing
combinations left, i.e. until $\text{Rank }Z_k \approx m_k$, or
until we establish a pattern if there are infinite $Z_k$'s. For
theories with finite reducibility level $L$, we can express the
number of independent constraints $m$ as
\begin{equation}\label{eq:cs3_graded_sum_m_i}
m = \sum_{i=0}^{L} (-)^i m_i.
\end{equation}
We introduce conjugate ghost pairs for every $Z_k$ as follows
\begin{eqnarray}
\eta^{a_k}, & \eps(\eta^{a_k})= \eps_{a_0} +k+1, & \text{gh }
\eta^{a_k} =1 + k\\
\PP_{a_k}, & \eps(\eta^{a_k})= \eps_{a_0} +k+1, & \text{gh }
\PP_{a_k}=-1-k,
\end{eqnarray}
where $\eps$ describes the Grassmann number and `gh', the ghost
number. Also, the Poisson bracket is as usual
\begin{equation}
[\PP_{a_k} , \eta^{b_j}] = (-)^{\eps_{a_0}+k}
\de^{b_k}_{a_k}\de_{kj}.
\end{equation}

We finally define the boundary terms in the ghost number one,
fermionic BRST charge
\begin{equation}\label{eq:cs3_Om_reducible}
Q = \eta^{a_0} g_{a_0} + \sum_{k=1}^L{(
\eta^{a_k}{(Z_k)_{a_k}}^{a_{k-1}}\PP_{a_k-1})} + \text{`more'},
\end{equation}
where it can be shown that the requirement $[Q,Q]=0$
determines the rest of the terms, and that $Q$ is unique
up to a canonical transformation.

\subsection{Construction of the pure spinor BRST operator}
As shown in appendix \ref{sec:sp7_U5}, the 16 component, pure
spinor $\la^\al$ possesses exactly 11 independent complex degrees
of freedom. This means that of the 10, complex, pure spinor
constraints, only 5 are independent and ghost for ghost terms are
needed in $Q_{gc}$. Information required to build $Q_{gc}$ to
level 4 is summarized in table \ref{tab:gc4_Z_identities} below.

\begin{table}[h]
\caption{Reducibility of pure spinor
constraints}\label{tab:gc4_Z_identities}
\begin{tabular}{|c|c|c|c|c|c|c|}
\hline\\
Level $k$ & $Z_k$ & Rank $Z_k$& $m_k$& ghosts &
$gh_{gc}$& $\eps$\\
\hline\\
0 & $\la\ga^m\la$ & 5 & 10 & $C_m,B^m$ & $1,-1$ & 1\\
1 & $(\ga_m\la)_\al$ & 5 & 16 &
$U^\al , V_\al$ & $2,-2$ & 0\\
2 & $(\la\ga_n\ga_p)^\al$ & 11 & 46 & $C^{np} , B_{np}$ & $3,-3$ &
1\\
3 & ${(\la\ga^{[n})_\be \de^{p]}_q + \frac{1}{10}(\la\ga_q)_\be
\eta^{np}}$& 35 & 160 & $U^{\be q}, V_{\be q}$ & $4,-4$ & 0\\
4 & $(\la\ga^{rs})^\be \eta^{tq}$ & 125 & 450 & $C_{rst}, B^{rst}$
& $5,-5$ & 1\\
4 & $\de^\be_{(\eta}(\la\ga^q)_{\ga)}$ & n/a & 136 & $C^{\eta\ga},
B_{\eta\ga}$ & $5,-5$ & 1\\ \hline \end{tabular}
\end{table}
We define $k$, $m_k$ and $Z_k$ as in section
\ref{sec:cs3_reducible_BRST}, where $Z_0$ corresponds to the pure
spinor constraints. The rank of the reducibility matrix $Z_k$ is
calculated on-shell and $Z_k$ must be chosen such that $\text{Rank
}Z_k$ is exactly equal to the number of redundant linear
combinations contained in $Z_{k-1}$ as in equation
\ref{eq:gc6_rank}. Ghosts denoted by $(C,B)$ are fermionic
conjugate pairs and those by $(U,V)$ are bosonic. We define $\eps$
to be the Grassmann parity of the ghosts and $gh_{gc}$ the ghost
constraint ghost number.

As there appears to be no obvious analytical way of deriving
ranks, they were calculated numerically for particular pure values
of $\la^\al$, using the computer software package Maple. It is
expected that the ranks remain constant for all pure values of
$\la^\al$, though no proof of this is provided here.

There are some further subtleties here. Notice that $C^{np}$ has
46 components, since it consists of an antisymmetric piece, with
45 components, and a trace piece with just one component. Note
also that $Z_3$ consists of the sum of the antisymmetric and trace
piece, with respect to $n,p$, of $(\la\ga^n)_\be \de^p_q$.

Below are the reducibility identities, the first of which
\eqref{eq:gc3_Z_1} completely describes the redundancy in the
constraints as in equation \eqref{eq:gc6_Zg}. Successive
identities take the form $Z_k Z_{k-1}\preprox 0$ as in
\eqref{eq:gc6_ZZ_0}. The complete equations for $Z_4 Z_3\preprox
0$ consist of two identities \eqref{eq:gc6_Z4Z3_1} and
\eqref{eq:gc6_Z4Z3_2}, whereas the rest require just one equation.
The gamma matrix identities in appendix \ref{sec:gamma_matrices}
may be used to confirm them.
\begin{eqnarray} \label{eq:gc3_Z_1}
(\ga_m\la)_\al \la\ga^m\la = 0, \\
(\la\ga_n \ga_p)^\al (\la\ga_m)_\al = \la\ga_m\la\eta_{np} +
\la\ga_n\la\eta_{pm} - \la\ga_p\la\eta_{nm} \preprox 0,\\
{[(\la\ga^{[n})_\be \de^{p]}_q + \frac{1}{10}(\la\ga_q)_\be
\eta^{np}]} (\la\ga_n\ga_p)_\al = -\half \la\ga^n\la(\ga_n\ga_p)_\be^\al \preprox 0,\\
{[(\la\ga^{rs})^\be \eta^{tq}]} {[(\la\ga^{[n})_\be \de^{p]}_q +
\frac{1}{10}(\la\ga_q)_\be
\eta^{np}]}  \preprox 0, \label{eq:gc6_Z4Z3_1}\\
{[\de^\be_{(\eta} (\la\ga^q)_{\ga)}]} {[(\la\ga^{[n})_\be
\de^{p]}_q + \frac{1}{10}(\la\ga_q)_\be \eta^{np}]} \preprox 0.
\label{eq:gc6_Z4Z3_2}
\end{eqnarray}

Let us study the first level reducibility condition
\eqref{eq:gc3_Z_1}, in detail. We firstly confirm the reducibility
condition using the Fierz identity of equation
\eqref{eq:B_Fierz_Clifford}. There are $m_1=16$ linear
combinations of $\la\ga^m\la$, denoted by $Z_1 = (\ga_m\la)_\al$,
which disappear globally. We calculate, using Maple, that
$\text{Rank }(\ga_m \la)_\al \preprox 5$, so there are only 5
linearly independent combinations, which match the 5 redundant
constraints in $\la\ga^m\la$. Therefore, $Z_1=(\ga_m \la)_\al$
contains $16-5=11$ redundant linear combinations of $\la\ga^m\la$,
which need to be taken care of at the next level.

If $\Qh_{gc}$ were to terminate at finite level $L$, we could
count the number of independent first class ghost constraints by
the graded sum in equation \eqref{eq:cs3_graded_sum_m_i}, with
$m=5$. This has an important bearing on the vanishing of the
central charge for the superstring, which is mentioned in section
\ref{sec:gc7_quant_open_string}. In the case of infinite ghosts
for ghosts, then the sum would need to be regularized.

The process of finding $Z_k$'s is largely a matter of trial and
error. Candidate $Z_k$'s are put forward which annihilate
$Z_{k-1}$, then their ranks are checked using Maple. We restrict
the search to $Z_k$'s linear in $\la^\al$, since higher order
powers of $\la^\al$ tend to have significantly higher $m_k$, i.e
redundancy, for a given rank.

Finally, we construct the ghost constraint, BRST operator, up to
level 2, in the manner of equation \eqref{eq:cs3_Om_reducible}.
\begin{equation}\label{eq:gc3_Q_pre_sp}
\begin{split}
\Qh_{gc}=& \Ch_m \lah\ga^m\lah + \Uh^\al (\lah\ga_m)_\al \Bh^m +
\Ch^{np}(\lah\ga_n\ga_p)^\al \Vh_\al + \ldots\\
&+ (\Bh_n \Bh_p \Ch^{np} -\half \Bh_m \Bh^m \Ch^{np}\eta_{np}) +
\ldots ,
\end{split}
\end{equation}
where the expressions in the first line are the boundary terms.
The anti-hermitian, ghost constraint ghost number operator is
given by
\begin{equation}\label{eq:gc4_G_pre}
\Gh_{gc} = \frac{i}{2}(\Ch_m \Bh^m - \Bh^m \Ch_m + 2 \Uh^\al
\Vh_\al + 2\Vh_\al\Uh^\al + 3\Ch^{np}\Bh_{np} - 3\Bh_{np}\Ch^{np}
+ \ldots).
\end{equation}


\section{Towards the covariant quantization of the $D=10$, $N=1$
superparticle} \label{sec:gc4_superp}

We proceed to build a quantum system in the Schr\"odinger
representation, in the manner of section \ref{sec:gc5_formal}, for
the superparticle theory with Berkovits BRST operator $\Qh =
\lah^\al \deeh_\al$, and with first class ghost constraints
described by $\Qh_{gc}$ in equation \eqref{eq:gc3_Q_pre_sp}.

The fact that $\Qh_{gc}$ is incomplete means that we cannot
explicitly calculate the BRST extension with respect to
$\Qh_{gc}$, of arbitrary operators which are gauge invariant with
respect to $\lah\ga^m\lah$. We tackle this issue in section
\ref{sec:gc7_pure_spinor_coho}, by constructing a basis for ghost
number zero operators in $H^0_{op}(\Qh_{gc})$, whose properties
can be deduced without having to build their respective BRST
extensions explicitly. The price to be paid is that we have only
one representative of each cohomology class of
$H^0_{op}(\Qh_{gc})$.

Our approach is systematic. We build the defining operators in
section \ref{sec:gc7_defining_operators}, the gauge-fixed action
in \ref{sec:gc7_BRST_sp_action}, the physical states in
\ref{sec:gc7_physical_states} and the physical operators in
\ref{sec:gc7_physical_operators}.

In the latter two sections, we begin with the ghost constraint
cohomology $H(\Qh_{gc})$ in subsections
\ref{sec:gc7_Q_pre_state_coho} and \ref{sec:gc7_pure_spinor_coho},
before the physical cohomology $H(\Qh|H(\Qh_{gc}))$ in
\ref{sec:gc7_phys_state_coho} and
\ref{sec:gc7_physical_operator_coho}, and we compare with the
Brink-Schwarz model in \ref{sec:gc7_BS_state_comparison} and
\ref{sec:gc7_BS_operator_comparison}.

We also construct the super-Poincar\'e covariant, inner product in
section \ref{sec:gc7_phys_state_coho} and draw an analogy with the
Witten, particle wavefunction for Chern-Simons theory in
\ref{sec:gc7_Chern_Simons}.

We further discover in subsection
\ref{sec:gc7_physical_operator_coho} that $\Qh$ indirectly implies
what we name as `effective constraints'. The operator cohomology
$H^0_{op}(\Qh|H^0_{op}(\Qh_{gc}))$ modulo these `effective
constraints' seems to correspond with the space of light-cone
gauge operators of the Brink-Schwarz model.

There are some useful, relevant results contained in appendix
\ref{sec:BRST_Schro} concerning BRST quantization in the
Schr\"odinger representation.


\subsection{The defining operators $\Qh_{gc}$, $\Gh_{gc}$, $\Qh$
and $\Gh$} \label{sec:gc7_defining_operators}

As already observed,
\begin{equation}\label{eq:gc5_nilpotence}
{\Qh}^2 = [-i\Bh^m \Ph_m , \Qh_{gc}] \presim 0,
\end{equation}
and we have the first terms of the anti-hermitian $\Gh_{gc}$ in
equation \eqref{eq:gc4_G_pre}. However, the ghost number operator
$i\lah^\al \wh_\al$ is gauge invariant
\begin{equation}
[i\lah^\al \wh_\al, \lah\ga^m\lah] \preprox 0,
\end{equation}
but requires a BRST extension in order to make it $Q_{gc}$-closed.
Working up to reducibility level 2 again, we find
\begin{eqnarray}\label{eq:gc6_G_prime}
\Gh' = i(\lah^\al \wh_\al - 2\Ch_m\Bh^m -3\Uh^\al\Vh_\al -
4\Ch^{np}\Bh_{np} - \ldots), & [\Gh', \Qh_{gc}] = 0.
\end{eqnarray}
Similarly, the anti-hermitian ghost number is given by
\begin{equation}
\begin{split}
\Gh =& \frac{i}{2}[\lah^\al \wh_\al +  \wh_\al\lah^\al -
2(\Ch_m\Bh^m - \Bh^m\Ch_m) -3(\Uh^\al\Vh_\al + \Vh_\al\Uh^\al) \\
&- 4(\Ch^{np}\Bh_{np} - \Bh_{np}\Ch^{np}) \ldots].
\end{split}
\end{equation}
Notice also
\begin{equation}\label{eq:gc7_G_from_G_pre}
\Gh = \frac{i}{2}[\lah^\al \wh_\al +  \wh_\al\lah^\al -
(\Ch_m\Bh^m - \Bh^m\Ch_m) - (\Uh^\al\Vh_\al + \Vh_\al\Uh^\al) -
\ldots] - \Gh_{gc}.
\end{equation}
Careful calculation reveals that the above relation holds for all
levels, at least if all $Z_k$ are linear in $\la^\al$.

We now have $\Qh_{gc}$, $\Gh_{gc}$, $\Qh$ and $\Gh$, which can be
confirmed to obey equations \eqref{eq:gc5_Q_pre_squared} to
\eqref{eq:gc5_ghost_op_commute} as required.

\subsection{The gauge-fixed, BRST invariant, superparticle
action}\label{sec:gc7_BRST_sp_action}
Given the action in equation \eqref{eq:gc4_S_intermediate}, it
remains to gauge-fix the pure spinor gauge symmetry in order to
obtain the full BRST action. If one chooses the gauge $\La_m=0$,
the full action is given by
\begin{equation}\label{eq:gc6_S_sp_full}
\begin{split}
S=\int{d\tau\, (\dot{X}^m P_m + \dot{\theta}^\al p_\al -
\frac{1}{2}P_m P^m + \dot{\la}^\al w_\al + }\\
\dot{C}_m B^m + \dot{U}^\al V_\al + \dot{C}^{np} B_{np} +\ldots).
\end{split}
\end{equation}
It is simplest to think of the gauge-fixing procedure from the
Hamiltonian point of view, where this corresponds simply to the
choice of zero gauge-fixing fermion and hence zero ghost
Hamiltonian. The intermediate `first class' Hamiltonian, by which
one means first class with respect to the ghost constraints, is
given by $H = 1/2 P_m P^m$, which is already BRST invariant with
respect to both $Q_{gc}$ and $Q$.

\subsection{The physical states}\label{sec:gc7_physical_states}

\subsubsection{The pure spinor, state cohomology $H^{\pm
k}_{st}(\Qh_{gc})$} \label{sec:gc7_Q_pre_state_coho}

A physical state obeys equations \eqref{eq:gc5_phys_state} and
\eqref{eq:gc5_phys_state2}, belonging to $H_{st}(\Qh |
H_{st}(\Qh_{gc}))$. A preliminary step is to obtain wavefunctions
in the usual Schr\"odinger representation, belonging to the state
cohomology $H^{\pm k }_{st}(\Qh_{gc})$. The ghost constraint ghost
numbers $\pm k$, which are undetermined since $\Qh_{gc}$ hasn't
yet been completed, refer to the cohomologies at which the pure
spinor constraints are imposed as Dirac constraints.

We specify states $\phi_{C=0,U=0}$ and $\phi_{B=0,V=0}$, which are
defined up to a normalization factor by $\Ch^m
\phi_{C=0,U=0}=\Uh^\al\phi_{C=0,U=0} = \ldots =0$ for all $\Ch$'s
and $\Uh$'s and similarly for $\phi_{B=0,V=0}$, where
\begin{eqnarray}
\Gh_{gc} \phi_{C=0,U=0} = k \phi_{C=0,U=0}, & \Gh_{gc}
\phi_{B=0,V=0} = -k \phi_{B=0,V=0}.
\end{eqnarray}
The wavefunctions are normalized as
\begin{eqnarray}\label{eq:gc7_phi_C_U}
\phi_{C=0,U=0} = (\prod_{m}{C_m}) \de^{(16)}(U)\ldots, &&
\phi_{B=0,V=0} =1,
\end{eqnarray}
so that $(\phi_{B=0,V=0},\phi_{C=0,U=0})=1$.

A wavefunction $\psi_{-}(\la) \phi_{B=0,V=0}$ in
$H^{-k}_{st}(\Qh_{gc})$ obeys,
\begin{eqnarray}\label{eq:gc6_psi_minus_Q_pre_closed} \Qh_{gc}
\psi_{-}(\la) \phi_{B=0,V=0} = 0 &\Imp& \lah\ga^m\lah
\psi_{-}(\la) =0
\end{eqnarray}
and there are no $\Qh_{gc}$-exact states at this ghost constraint
ghost number. On the other hand, a wavefunction
$\psi_{+}(\la)\phi_{C=0,U=0}$, in the isomorphic state cohomology
$H^{k}_{st}(\Qh_{gc})$ is $\Qh_{gc}$-closed for any function
$\psi_{+}(\la)$, but we can vary the wavefunction by arbitrary
$\Qh_{gc}$-exact amounts
\begin{eqnarray}
\de \psi_+(\la)\phi_{C=0,U=0} &=& i\Qh_{gc}\, f_m(\la) \Bh^m
\phi_{C=0,U=0}\\
\Imp \de\psi_+(\la) &=& f_m(\la) \la\ga^m\la,
\end{eqnarray}
for arbitrary wavefunction $f_m(\la)$. As usual, the two
cohomologies are isomorphic
\begin{equation}
H^{k}_{st}(\Qh_{gc}) \cong H^{-k}_{st}(\Qh_{gc}).
\end{equation}

Note, the fact that there are ghosts for ghosts doesn't affect the
state cohomology. It only becomes important in the operator
cohomology, where they are there to cancel the 5 excess degrees of
freedom hidden in the 10 ghosts $C_m$.

See appendix \ref{sec:BRST_Schro} for a brief discussion of the
BRST concepts which have arisen in this subsection.

\subsubsection{The physical state cohomologies
$H^{\pm(g+1)}_{st}(\Qh|H_{st}^{\pm k}(\Qh_{gc}))$}
\label{sec:gc7_phys_state_coho}

In the previous subsection, we have not yet worried about
requiring $\Qh \psi \presim 0$ or constraining the generic
wavefunctions $\psi_{\pm}(\la)$ to a particular ghost number. We
define the ghost part of generic physical wavefunctions as
\begin{eqnarray}
\phi_g = \phi_{w=0, C=0, U=0}, && \phi_{-g} = \phi_{\la=0, B=0,
V=0}
\end{eqnarray}
where,
\begin{eqnarray}
\phi_g \in H^{k}_{st}(\Qh_{gc}), && \phi_{-g} \in H^{-k}_{st}(\Qh_{gc}),\\
\Gh \phi_{\pm g} = \pm g \phi_{\pm g}, && g=\frac{11}{2} - k,
\end{eqnarray}
where the expression for $g$ in terms of $k$ is deduced from
\eqref{eq:gc7_G_from_G_pre} and \eqref{eq:cs3_graded_sum_m_i}. The
wavefunctions are normalized as in equation
\eqref{eq:gc7_phi_C_U}, thus $(\phi_{-g},\phi_{g})=1$. There are
two ingredients to each of the states $\phi_g$ and $\phi_{-g}$. In
$\phi_{-g}$ for example, there is firstly a delta-function, which
fixes 5 components of $\la^\al$ in terms of the other 11, so that
equation \eqref{eq:gc6_psi_minus_Q_pre_closed} is obeyed.
Secondly, there is a delta function to set the remaining 11
components of $\la^\al$ to zero, in order to provide the standard
ghost number $g$ state. The combined wavefunction $\phi_{\la=0}$
is straightforward, however it seems difficult to split it into
the two aforementioned parts. The generic ghost number $(g+1)$
wavefunction is given by
\begin{equation}
\psi_{g+1} = \lah^\al A_\al(X,\theta) \phi_g,
\end{equation}
which is simply our version of the Berkovits ghost number one
wavefunction. The conditions of BRST invariance
\eqref{eq:gc5_phys_state} imply, in a similar manner to Berkovits
in section \ref{sec:sp6_berk_superp}, that $A_{\al}(X,\theta)$
obeys the super-Maxwell equations of motion
\begin{equation}
\ga_{mnpqr}^{\al\be}D_\al A_\be = 0,
\end{equation}
since $\lah^\al \lah^\be D_\al A_{\be}(X,\theta)\phi_g \presim 0$,
and $\ga_m^{\al\be}\lah\ga^m\lah D_\al A_\be\phi_g$ is
$\Qh_{gc}$-exact. Also, $\psi_{g+1}$ is $\Qh_{gc}$-closed for
arbitrary $A_\al(X,\theta)$.

The BRST transformation of the wavefunction $\psi_{g+1}$ is
\begin{equation}
\de \psi_{g+1} = \Qh\, \La(X,\theta) \phi_g,
\end{equation}
where $\La(X,\theta) \phi_g$ is $\Qh_{gc}$-closed for arbitrary
$\La(X,\theta)$, which implies the usual super-Maxwell gauge
transformation
\begin{equation}
\de A_\al(X,\theta) = D_\al \La(X,\theta).
\end{equation}

Our wavefunction $\psi_{g+1}$ couples in the inner product to
certain states at opposite ghost numbers, as in equation
\eqref{eq:gcB_psi_g_psi_g_prime},  given by
\begin{equation}
\psi_{(-g-1)} = \wh_\al \At^\al(X,\theta) \phi_{-g},
\end{equation}
The conditions of BRST invariance equation
\eqref{eq:gc5_phys_state} imply the equation of motion
\begin{equation}\label{eq:gc6_At_eqn}
D_\al \At^\al(X,\theta) =0.
\end{equation}
The BRST transformation of the wavefunction $\psi_{(-g-1)}$ is
\begin{eqnarray}\label{eq:gc6_At_Q_closed}
\de \psi_{(-g-1)} = \Qh B^{\al\be}(X,\theta) \wh_{\al}\wh_{\be}
\phi_{-g} &\text{for}&\Qh_{gc} B^{\al\be}(X,\theta)
\wh_{\al}\wh_{\be} \phi_{-g}=0,
\end{eqnarray}
which implies the following gauge transformation
\begin{eqnarray}\label{eq:gc6_At_gauge_transf}
\de \At^\al(X,\theta) = D_{\be}B^{\al\be}(X,\theta) && \text{for
} \ga^m_{\al\be} B^{\al\be}=0.
\end{eqnarray}

We expect the two cohomologies to be isomorphic
\begin{equation} H^{(g+1)}_{st}(\Qh|
H_{st}^{k}(\Qh_{gc})) \cong H^{(-g-1)}_{st}(\Qh|
H_{st}^{-k}(\Qh_{gc})),
\end{equation}
and we relate them in section \ref{sec:gc7_BS_state_comparison}
through the Schr\"odinger inner product, which we now define.

Since $\la^\al$ and $w_\al$ are complex, we define
\begin{eqnarray}
\lab^\al = (\la^\al)^* , && \wb_\al = (w_\al)^*,
\end{eqnarray}
thus,
\begin{eqnarray}
(\lah^\al)^\dag = \labh^\al,  && (\wh_\al)^\dag = \wbh_\al.
\end{eqnarray}
Given a state $\psi$, we define $\psib =\psi^*$. Thus, in
particular
\begin{eqnarray}
\psib_{g+1} = \labh^\al A_{\al}(X,\theta)\phi_g, && \psib_{(-g-1)}
= \wbh_\al \At^\al(X,\theta) \phi_{-g},
\end{eqnarray}
where we choose the constant phase factors present in $\psi_{g+1}$
and $\psi_{(-g-1)}$ such that $A_{\al}(X,\theta)\phi_g$ and
$\At^\al(X,\theta) \phi_{-g}$ are both real. We find that by
replacing $\Qh_{gc}$ and $\Qh$ with $\Qh_{gc}^\dag$ and
$\Qh^{\dag}$ respectively, the condition that $\psib_{g+1}$ and
$\psib_{(-g-1)}$ be BRST closed
\begin{eqnarray}
\Qh_{gc}^\dag \psib_{g+1}=0, && \Qh^\dag \psib_{g+1} \presim 0,\\
\Qh_{gc}^\dag \psib_{(-g-1)}=0, && \Qh^\dag \psib_{(-g-1)} \presim
0, \label{eq:gc6_psib_m_g_m_one_Q_Q_pre_closed}
\end{eqnarray}
implies exactly the same equations of motion for $A_\al$ and
$\At^\al$ as before. Also, the BRST transformations of
$\psib_{g+1}$ and $\psib_{(-g-1)}$ imply exactly the same gauge
transformations of $A_\al$ and $\At^\al$.

In the inner product, we choose the convention of initially
placing $\psib_{(-g-1)}$ on the left hand side, though we could
just have easily chosen $\psib_{g+1}$. Crucially, a generic
BRST-exact operator, as in equation \eqref{eq:gc5_Qph_exact},
obeys
\begin{eqnarray}
(\psib_{(-g-1)}, \{[\Uh, \Qh] +[\Uh_{gc},\Qh_{gc}]\}\psi_{g+1}) =
0 && \text{for } [\Uh,\Qh_{gc}]=0,
\end{eqnarray}
using equations \eqref{eq:gc5_phys_state},
\eqref{eq:gc6_psib_m_g_m_one_Q_Q_pre_closed} and the Jacobi
identity. So we have seen that, by replacing $\psi$ with $\psib$
on the left of the inner product, the fact that $\Qh$ and
$\Qh_{gc}$ aren't hermitian isn't problematic.

Let us calculate a general inner product,
\begin{equation}\label{eq:gc6_psi_scalar_prod}
\begin{split}
(\psib_{(-g-1)},\psi_{g+1}) &= \int{[d^{10} X \,d^{16}\theta
\,d^{16}\la \,d^{16}\lab \,d^{10}C \,d^{16}U \ldots ]
(\psib_{(-g-1)})^* \psi_{g+1}}\\
&=\int{d^{10} X \,d^{16}\theta \, \At^\al(X,\theta)
A_{\al}(X,\theta)}.
\end{split}
\end{equation}
This tells us that on expanding $A_{\al}(X,\theta)$ in powers of
$\theta^\al$, the coefficient of the $(\theta)^i$ term in
$A_{\al}(X,\theta)$ couples to the coefficient of the
$\theta^{(16-i)}$ term in $\At^\al(X,\theta)$. Therefore just as
we expand $A_\al$ in increasing powers of $\theta^\al$ starting at
$1$ as in equation \eqref{eq:sp7_A_al_cmpts}, so it makes sense to
expand $\At^\al$ in decreasing powers of $\theta^\al$ starting
with $\theta^1\theta^2\ldots\theta^{16}$. For this purpose, we
invent a useful notation,
\begin{equation}
\tht_\al \tht_\be \ldots =
\frac{\pa}{\pa\theta^\al}\frac{\pa}{\pa\theta^\be}\ldots (\theta^1
\theta^2\ldots \theta^{16})
\end{equation}
where $\thh^\al$ denotes that $\theta^\al$ has been excluded from
the product and the sign depends on $\al,\be$ etc.. Also,
\begin{equation}
\theta^\rho \tht_\al \tht_\be\ldots = \frac{\pa}{\pa
\tht_{\rho}}\tht_\al \tht_\be \ldots,
\end{equation}
so that the covariant derivative can be written
\begin{equation}
D_\al \equiv \tht_{\al} - i \ga^m_{\al\be}\pa_m
\frac{\pa}{\pa\tht_\be},
\end{equation}
which is useful for making component calculations with $\At^\al$.
We observe that the gauge transformation of $\At^\al$ does a
similar job to the equation of motion for $A_\al$ and vice versa.
The superfield $\At^\al$ has two physical components, $\at^m(X)$
and $\chit_\al(X)$. We can choose a special gauge for $\At^\al$
analogous to that for $A_\al$ in equation
\eqref{eq:sp7_A_al_cmpts}, such that
\begin{equation}
\At^\al = i \at^m(X) \ga_m^{\al\be} \tht_\be - \chit_{\ga}(X)
\ga_m^{\al\be}\ga^{m\ga\de} \tht_\be \tht_\de + \ldots,
\end{equation}
and all remaining components depend only on $\at^m$ and
$\chit_\al$. The equation of motion \eqref{eq:gc6_At_eqn} implies
\begin{equation}
\pa_m \at^m(X) =0,
\end{equation}
and the gauge transformation \eqref{eq:gc6_At_gauge_transf}
implies
\begin{eqnarray}\label{eq:gc7_antifield_gauge_sym}
\de \at^m = \pa_n(\pa^n s^m - \pa^m s^n), && \de \chit_\al =
\ga^m_{\al\be}\pa_m \xi^\be,
\end{eqnarray}
for arbitrary parameters $s^n(X)$ and $\xi^\be(X)$. Notice that
the inner product, in the last line of equation
\eqref{eq:gc6_psi_scalar_prod}, is gauge invariant with respect to
variations in $A_\al$ due to the equations of motion of $\At^\al$,
and vice versa.

A derivation of the expression for \eqref{eq:gc6_psi_scalar_prod}
in terms of component fields has yet to be completed, due to the
length of the calculation. Nevertheless, we deduce that up to
normalization factors
\begin{equation}\label{eq:gc6_scalar_prod_cpts}
(\psib_{(-g-1)},\psi_{g+1}) = \int{d^{10}X \, (\at^m(X)a_m(X) +
\chit_\al(X) \chi^\al(X))},
\end{equation}
for a number of reasons. Firstly, it must be gauge invariant and
thus depend only on physical components. Secondly, since $A_\al$
and $\At^\al$ are linear in their physical components, the inner
product must also be linear in them. Thirdly, since $a_m$ appears
only at odd powers of $\theta^\al$, and $\chi^\al$ only at even
powers of $\theta^\al$ in $A_\al$, and since $\at^m$ appears only
at odd powers of $\tht_\al$ and $\chit_\al$ only at even powers of
$\tht_\al$ in $\At^\al$, so the inner product must be a sum of
just two terms, one dependent only on $a_m$ and $\at^m$, the other
only on $\chi^\al$ and $\chit_\al$. The above expression is the
only gauge invariant possibility which fits the above criteria.

In order to write down a basis for physical states $\psi_{g+1}$,
we create states with definite quantum numbers, which are defined
up to a BRST-exact wavefunction
\begin{eqnarray}
\psi_{g+1} \sim \psi_{g+1}(k_m, a_m , \chi^\al),
\end{eqnarray}
where $k_m$, $a_m$ and $\chi^\al$ are all constant, real numbers,
such that
\begin{eqnarray}
k^2=0, & k^m a_m=0,  & \ga^m_{\al\be}k_m \chi^\be=0.
\end{eqnarray}
Similarly,
\begin{eqnarray}
\psi_{(-g-1)} \sim \psi_{(-g-1)}(k_m, \at^m , \chit_\al),
\end{eqnarray}
where $k_m$, $\at^m$ and $\chit_\al$ are all constant, real
numbers, such that
\begin{eqnarray}
k^2=0, && k_m \at^m=0.
\end{eqnarray}

\subsubsection{Comparison with the light-cone gauge, BS
superparticle}\label{sec:gc7_BS_state_comparison}

We relate a state $\psi_{g+1}(k_m, a_m , \chi^\al)$ to its light
cone gauge, BS equivalent, by gauge-fixing $a^+=0$, as in appendix
\ref{sec:sp3_S_S_Y_M}. This will be important for comparing
operators of the BS superparticle in the light-cone gauge with
operators in our superparticle, by observing how they act on
equivalent states. We write a generic wavefunction $\psi_{g+1}$
with light-cone gauge values of $k_m$, $a_m$ and $\chi^\al$
\begin{equation}\label{eq:gc6_psi_gplusone_LC}
\psi_{g+1} \presim \psi^{LC}_{g+1}(k_m , a_m , \chi^\al )
\end{equation}
where $a^-(a^i) = (k^+)^{-1}k^i a^i$, and where we are using the
standard light-cone gauge notation as in appendix
\ref{sec:gc2_review}. Also, $\chi^b$ is determined as a function
of $\chi^{\bd}$ as in equation \eqref{eq:sp3_chi_a_from_chi_adot},
since $\chi^\al$ obeys the Dirac equation. This maps directly to
the semi-light-cone gauge Brink-Schwarz wavefunction $\psi_{BS}$,
with the usual notation as described in appendix
\ref{sec:sp2_superp}.
\begin{equation}\label{eq:gc6_psi_BS}
\psi_{BS} = \exp{(ik_mX^m)}(a^i|i> +
-i2^{\frac{1}{4}}(P^+)^{-\half}\chi^{\bd}|\bd>).
\end{equation}
Likewise,
\begin{equation}
\psib_{(-g-1)} \presim \psib^{LC}_{(-g-1)}(k_m,\at^m, \chit_\al),
\end{equation}
where we fix the gauge symmetry of equation
\eqref{eq:gc7_antifield_gauge_sym} with conditions $\at^+=0$ and
$\chit^b = 0$.

We compare our Schr\"odinger inner product to that of the
semi-light cone gauge, BS superparticle. From equation
\eqref{eq:gc6_scalar_prod_cpts}, we learn
\begin{equation}
(\psib_{(-g-1)}^{LC}, \psi_{g+1}^{LC} ) = \int{d^{10}X \,(\at^i
a^i + \chit^{\bd} \chi^{\bd})},
\end{equation}
which agrees with the semi-light-cone gauge inner product in
equation \eqref{eq:sp2_BS_sc_prod} up to a normalization factor.
We therefore make the map between the two, isomorphic, state
cohomologies
\begin{equation}
\psi^{LC}_{g+1}(k_m
,(0,a^-(a^i),a^i),(\chi^b(\chi^{\bd}),\chi^{\bd})
\longleftrightarrow \psib^{LC}_{(-g-1)}(k_m
,(0,\at^-(\at^i),\at^i),(0,\chit^{\bd})),
\end{equation}
in a similar manner to equation \eqref{eq:gcB_coho_map}.

\subsubsection{An analogy with abelian Chern-Simons theory}
\label{sec:gc7_Chern_Simons}

The manner in which the physical wavefunction obtained from the
Berkovits BRST operator describes super Yang-Mills is unusual. In
particular, the wavefunction $\la^\al A_\al$ appears at a ghost
number one higher than that required to impose the constraints in
$\Qh$ as Dirac constraints. It was noticed however
\cite{Berkovits:2001rb}, that there is a more simple precedent.
Witten \cite{Witten:1992fb} shows how Chern-Simons theory arises
in a similar way from a string theory, which is easily modified to
a particle theory. An analogy can be drawn between the Witten
particle theory and the Berkovits superparticle theory. Our
analogy differs significantly to that of Berkovits'
\cite{Berkovits:2001rb}.

Abelian, Chern-Simons theory can be described by the world-line,
Witten action
\begin{equation}
S=\int{d\tau \, (\Xd^m P_m - l^m P_m)},
\end{equation}
where $m=0,1,2$, which is the Hamiltonian form of the theory
described by a zero Lagrangian. First class constraints $P_m$
imply the BRST operator
\begin{equation}
Q = -ic^m \pa_m,
\end{equation}
where $c^m ,b_m$ are conjugate pairs of fermionic ghosts. The most
general wavefunction can be expressed
\begin{equation}\label{eq:sp6_chern_simons_psi}
\psi_W = C(X) + c^m A_m(X) + \frac{i}{2}c^m c^n
\eps_{mnp}A^{*p}(X) + \frac{i}{6}c^m c^n c^p \eps_{mnp} C^*(X),
\end{equation}
which terminates because $c^m$ is fermionic. The condition
$Q\psi=0$, together with the BRST transformation $\de\psi = Q
\Om(c,X)$ for the particle model, imply the equations of motion
and gauge transformations for the Chern-Simons fields
\begin{eqnarray}
\pa_{[m}A_{n]} = 0, && \de A_m = \pa_m \La,\\
\pa_p A^{*p} = 0, && \de A^{*p} = \eps^{pmn}\pa_m w_n,\\
\pa_m C = 0, && \de C^*  = \pa_p u^p.
\end{eqnarray}
The Chern-Simons action is given by
\begin{equation}
\int{d^3 X \, (\half \eps^{mnp} A_m \pa_n A_p + i A^{*p} \pa_p
C)},
\end{equation}
where $A^{*p}$ is the antifield to $A_p$ and $C$ and $C^*$ are the
ghost and anti-ghost, and can be written remarkably compactly as
\begin{equation}\label{eq:sp6_ch_si_S_psi}
S = \half (\psi\, , Q \,\psi),
\end{equation}
where the inner product measure is the usual $d^3 X d^3 c$.

To make the analogy clear, we rewrite the Witten wavefunction for
the corresponding particle theory in the form
\begin{equation}
\psi_W = C(X)\phi_{b=0} + A_m(X)\ch^m \phi_{b=0} + i
A^{*m}(X)\bh_m\phi_{c=0} + i C^*(X)\phi_{c=0},
\end{equation}
where $\Gh\phi_{b=0} = -3/2 \phi_{b=0}$ and $\Gh\phi_{c=0} = 3/2
\phi_{c=0}$. The point to notice is that fields couple to
anti-fields in the Schr\"odinger inner product $(\psi_W,\psi_W)$,
since $(\phi_{b=0}, \phi_{c=0})=1$ and $(\ch^m \phi_{b=0} , \bh_n
\phi_{c=0}) = -i\de^m_n $, and all other inner products are zero.
In other words, states at opposite ghost number, as described by
anti-hermitian ghost number operator $\Gh$, couple to each other,
as in equation \eqref{eq:gcB_psi_g_psi_g_prime}. Furthermore, the
state cohomology $H^g_{st}(\Qh)$ is dual to $H^{-g}_{st}(\Qh)$,
and as part of their special relation, the equation stating that
$\psi_g \in H^g_{st}(\Qh)$ is BRST-closed is connected in a
specific way to the equation that states the BRST transformation
for $\psi_{-g} \in H^{-g}_{st}(\Qh)$, and vice versa. This is the
reason why fields and anti-fields appear at opposite ghost number,
since the equations of motion of a field are related in just the
right way to the gauge invariances of the corresponding antifield,
and vice versa.

Our wavefunction for the superparticle is thus
\begin{equation}
\psi = C(X,\theta) \phi_g + A_\al(X,\theta)\lah^\al\phi_g +
A^{*\al}(X,\theta)\wh_\al \phi_{-g} +  C^*(X,\theta) \phi_{-g},
\end{equation}
where $A^{*\al}(X,\theta)$ is the super-antifield to
$A_\al(X,\theta)$ and $C^*(X,\theta)$ the super-antighost to
super-ghost $C(X,\theta)$ and where we recall that $\phi_g =
\phi_{w=0,\ldots}$ and $\phi_{-g} = \phi_{\la=0,\ldots}$.

We might expect to be able to write a BV, superspace action for
super Maxwell in an analogous way to the Chern-Simons action in
equation \eqref{eq:sp6_ch_si_S_psi}, however the action
\begin{equation}
S= (\psib, \Qh \psi) = \int{d^{10}X d^{16}\theta\,
A^{*\al}(X,\theta)D_\al C(X,\theta)},
\end{equation}
unfortunately only provides the BV super-ghost, super-anti-field
part. This is because we need a ghost number $-(g+2)$ term in
$\psi$ in order to couple to the term $\Qh
A_\al(X,\theta)\lah^\al\phi_g$. Of course this should come as no
surprise, since there is no known action principle for the
superspace formulation of $D=10$ super Yang-Mills. The superfield
ghosts and anti-field are only realized on-shell.

A different analogy to Chern-Simons theory is also drawn by
Berkovits\cite{Berkovits:2001rb}. The main difference is that he
chooses a particular non-linear measure, similar to that used in
his expression for massless tree-level amplitudes, instead of the
natural measure used here. The principle is still the same, that
$C$ must couple to $C^*$, and $A$ to $A^*$ under the inner
product. However, due to the form of the measure, $A^*$ and $C^*$
appear at ghost numbers $g+2$ and $g+3$ respectively, though with
extra indices i.e $A^{*\al\be}$ and $C^{*\al\be\ga}$. It is
claimed however that their on-shell physical components still
correspond with the BV anti-fields of super-Maxwell.

\subsection{The physical operators}
\label{sec:gc7_physical_operators}
\subsubsection{The pure spinor, operator cohomology
$H_{op}^{0}(\Qh_{gc})$} \label{sec:gc7_pure_spinor_coho}

Due to the fact that no completion of $\Qh_{gc}$ has been found
yet, we cannot calculate all operators belonging to
$H_{op}^{0}(\Qh_{gc})$ explicitly. However, if we further restrict
ourselves to operators which have zero ghost numbers, as do
physical operators, we can specify a basis for all cohomology
classes. Furthermore we can deduce the algebra of the basis
elements, which is closed, and how the basis elements act on
physical states.

The operators $\lah^\al \wh_\al$ and $\lah^\al
{{\ga^{mn}}_\al}^\be \wh_\be/2$ form a basis for ghost number zero
operators, which are gauge invariant with respect to the first
class, ghost constraints. A general such gauge invariant operator
is of the form
\begin{equation}\label{gc7:F_gh0_gi}
\Fh = :\Fh_0(\lah^\al \wh_\al, \half \lah\ga^{mn}\wh): +
:\Fh_{1m}(\lah,\wh) \lah\ga^m\lah:,
\end{equation}
for some convenient normal ordering, where $F_{1m}$ has ghost
number $-2$, but is otherwise arbitrary and where the second term
vanishes on the constraint surface.

The BRST extensions of the basis elements are $\lah^\al \wh_\al
+\Eh$ and $\lah\ga^{mn}\wh/2 + \Lh^{mn}$, where
\begin{eqnarray}
\Eh=  2\Bh^m \Ch_m -3\Uh^\al\Vh_\al + 4\Bh_{np}\Ch^{np} - \ldots,\\
\Lh^{mn} =\Ch^m \Bh^n - \Ch^n \Bh^m + \half\Uh^\al
{(\ga^{mn})_\al}^\be \Vh_\be + \ldots.
\end{eqnarray}
We notice that $\lah\ga^{mn}\wh/2$ generates Lorentz
transformations for $\lah^\al$ and $\wh_\al$, and hence the
equation $[\lah\ga^{mn}\wh/2 + \Lh^{mn}, \Qh_{gc}]=0$ implies that
$\Lh^{mn}$ is the Lorentz generator for the ghost constraint
ghosts $\Ch$, $\Bh$, ..., if we assume $\Qh_{gc}$ to be a Lorentz
scalar. Therefore, the algebra of $\Eh$ and $\Lh^{mn}$ is given by
\begin{eqnarray}
[\Eh, \Lh^{mn}]=0 & [\Lh^{mn},\Lh^{pq}] = i(\eta^{np}\Lh^{mq} -
\eta^{mp}\Lh^{nq} + \eta^{mq}\Lh^{np} - \eta^{nq}\Lh^{mp}).
\end{eqnarray}

The scheme for constructing the BRST extension of any gauge
invariant operator, is firstly to split it into the form of
equation \eqref{gc7:F_gh0_gi} and discard the piece proportional
to $\lah\ga^m\lah$, whose BRST extension is always
$\Qh_{gc}$-exact. We then replace $\lah^\al \wh_\al$ and $\lah^\al
{{\ga^{mn}}_\al}^\be \wh_\be/2$ in $\Fh_0$ with their BRST
extensions $(\lah^\al \wh_\al + \Eh)$ and $(\lah^\al
{{\ga^{mn}}_\al}^\be \wh_\be/2 + \Lh^{mn})$.

We deduce that $\Lh^{mn}$ annihilates physical states due to being
antisymmetric in ghost constraint ghosts, and $\Eh \phi_{g}=0$,
but $\Eh \phi_{-g} = -i(5+k)\phi_{-g}$, where recall that $k$ is
an unknown constant. If necessary, the unknown constant $-i(5+k)$
can be subtracted from $\Eh$ to begin with, or when calculating
expectation values, we may place the negative ghost number
wavefunction on the left hand side of the inner product.

By following the above scheme, we can perform matrix element
calculations and compute quantum brackets of physical operators
etc.. The price to be paid is that, each cohomology class of
$H_{op}^{0}(\Qh_{gc})$ has only one representative using our
basis.

\subsubsection{The physical operator cohomology $H^0_{op}(\Qh|
H^0_{op}(\Qh_{gc}))$ modulo `effective constraints'}
\label{sec:gc7_physical_operator_coho}

A basis for operators belonging to $H^0_{op}(\Qh |
H^0_{op}(\Qh_{gc}))$, linear in phase-space variables, is given by
\begin{equation}\label{eq:gc7_basis_ops}
\Ph_m, \qh_\al, \Kh^{mn}, \Jh,
\end{equation}
where
\begin{equation}
\qh_\al=\ph_\al + i\Ph_m(\ga^m \thetah)_\al
\end{equation}
\begin{equation}\label{eq:gc7_K_mn}
\Kh^{mn}=\Xh^m \Ph^n - \Xh^n \Ph^m + \half \thetah^\al
{(\ga^{mn})_\al}^\be \ph_\be  + \half \lah^\al
{(\ga^{mn})_\al}^\be \wh_\be + \Lh^{mn},
\end{equation}
\begin{equation}
\Jh =2\Xh^m\Ph_m + \thetah^\al \ph_\al + \lah^\al \wh_\al + \Eh.
\end{equation}

Any $\Qh_{gc}$-closed, ghost number -1 operator $\Ah\in
H^0_{op}(\Qh_{gc})$ is also $\Qh_{gc}$-exact, because its gauge
invariant piece $\Ah_0(\lah,\wh)$ must be proportional to
$\lah\ga^m\lah$. Thus, $[\Ah,\Qh]\presim 0$, meaning that there
are no $\Qh$-exact operators, which aren't trivial, i.e
$\Qh_{gc}$-exact.

The operator cohomology $H^0_{op}(\Qh | H^0_{op}(\Qh_{gc}))$
cannot correspond with the light-cone gauge space of operators for
the BS superparticle for two reasons. There is no mass-shell
constraint $\Ph^2=0$, which would render $\Ph^2$ BRST exact, and
there are 8 too many independent fermionic operators $\qh_\al$,
compared to the 8 $\thetah^a$'s of the light-cone gauge BS
superparticle. However, something interesting happens, which saves
us. We find that the matrix element of $\Ph^2$ between arbitrary
physical states $(\psib_{(-g-1)},\Ph^2\psi_{g+1})$, which we
denote as $<\Ph^2>$, obeys,
\begin{equation}
<\Ph^2> = i(\psib_{(-g-1)} ,\, \Qh \pa^m A_m(X,\theta)\phi_g)=0,
\end{equation}
where we have used the super-Maxwell field equation $\pa^m
F_{m\al}=0$, where $F_{m\al}$ is the spin $3/2$ super
field-strength and $A_m$ the space-time super gauge connection, as
in appendix \ref{sec:sp3_S_Y_M_superspace}. This perhaps isn't so
surprising, since the striking feature of super-Maxwell in ten
dimensions is that the constraint equations alone place the theory
on-shell. Also,
\begin{equation}\label{eq:gc6_P_Q_al_eff_constraint}
<\Ph_m \ga^{m\al\be} \qh_{\be}> =-4(\psib_{(-g-1)} ,\, \Qh W^\al
\phi_g) =0
\end{equation}
where we use the abelian form of the constraint equation
\eqref{eq:sp3_Y_M_constraints}, the field equation $\pa^m
F_{m\al}=0$ and the identity $D_\be W^\al =
F_{mn}{(\ga^{mn})^\al}_\be W^\be/2$, where $W^\be =
\ga^{m\al\be}F_{m\al}$ is the photino superfield strength. Since
$\qh_\al$ obeys the Dirac equation in
\eqref{eq:gc6_P_Q_al_eff_constraint}, it effectively has the
required 8 independent degrees of freedom.

We describe $\Ph^2=0$ and $\Ph_m \ga^{m\al\be} \qh_{\be}=0$ as
`effective constraints', since they arise only indirectly from the
Berkovits BRST operator $\Qh$. All other `effective constraints'
are formed from these two expressions.

Given a generic effective constraint $\Gh_{\text{eff}}$, we deduce
that
\begin{eqnarray}\label{eq:gc6_eff_constraint_times_A}
<\Gh_{\text{eff}}\Ah> = <\Ah \Gh_{\text{eff}}> = 0 && \text{iff }
\Ah \in H^0_{op}(\Qh|H^0_{op}(\Qh_{gc})).
\end{eqnarray}
An interesting inference is that
\begin{eqnarray}\label{eq:gc7_eff_constr_first_class}
[\Ah,\Ph^2] \approx 0, && [\Ah, \Ph_m \ga^{m\al\be} \qh_{\be}]
\approx 0,
\end{eqnarray}
given $\Ah \in H^0_{op}(\Qh|H^0_{op}(\Qh_{gc}))$, where the
$\approx$ refers to the effective constraint surface. We observe
that the effective constraint surface is the same as the first
class part of the BS superparticle constraint surface, which
describes Siegel's superparticle model. We have now completed the
superparticle model, with the exception of not having produced an
explicit completion of the pure spinor BRST operator $\Qh_{gc}$,
which as argued in section \ref{sec:gc7_pure_spinor_coho}, is not
as restrictive as one might expect.

It seems plausible that $H^0_{op}(\Qh|H^0_{op}(\Qh_{gc}))$ modulo
the effective constraints, corresponds with the space of
light-cone gauge, BS operators. In the next subsection, we
explicitly construct the map from the light-cone gauge BS
operators to our `physical' operators.

\subsubsection{Comparison with the light-cone gauge, BS
superparticle}\label{sec:gc7_BS_operator_comparison}

Since we can map between any state $\psi_{g+1}(k_m,a_m,\chi^\al)$
in our model and the corresponding state
$\psi_{BS}(k_m,a^i,\chi^{\ad})$ in the light-cone gauge, BS
superparticle as in section \ref{sec:gc7_BS_state_comparison}, we
can also relate operators in the two models, which can be defined
by how they act on the states. We attempt to map the physical
operators of the light-cone gauge, BS superparticle to our basis
operators in equation \eqref{eq:gc7_basis_ops}.

Our $\Ph_m$'s, combined with effective constraint $<\Ph^2>=0$,
straightforwardly map to the $\Ph_m$'s of the BS model. The
fermionic operators $\qh_\al$ of our model, in the Schr\"odinger
representation, are simply the supersymmetry generators
\begin{equation}
Q_\al \equiv \frac{\pa}{\pa\theta^\al} + i
\ga^m_{\al\be}\theta^\be \pa_m.
\end{equation}
As a result of the effective constraint
\eqref{eq:gc6_P_Q_al_eff_constraint}, we can write $Q^{\ad}$ in
terms of $Q^a$,
\begin{equation}
<Q^{\ad}> = <2^{-\half} (\Ph^+)^{-1} \Ph^i \si^i_{\ad a}Q^a>,
\end{equation}
where $Q_\al = (Q^a, Q^{\ad})$, and $\si^i_{\ad a}$ are the
$SO(8)$ gamma matrices defined in appendix
\ref{sec:gamma_matrices}. Therefore, $Q^{\ad}$ is redundant.

Let us see how $Q^{a}$ behaves by observing how it acts on a
generic state $\psi_{g+1}$. We firstly choose a representative
physical state from each cohomology class, with light-cone gauge
quantum numbers as in section \ref{sec:gc7_BS_state_comparison}
\begin{eqnarray}
\psi_{g+1}^{LC}=\psi_{g+1}(k_m , (0,a^-(a^i),a^i) ,
(\chi^a(\chi^{\ad}),\chi^{\ad})),
\end{eqnarray}
and now calculate
\begin{equation}\label{eq:gc6_Q_al_psi}
Q_\al \psi_{g+1}((k_m , a_m , \chi^\be)) \sim \psi_{g+1}(k_m ,
(\ga_m\chi)_\al , - k_m a_n{(\ga^{mn})_\al}^\be).
\end{equation}
We can obtain this either with a calculation of $Q_\al A_\be$ in
components, or more simply, by reading off the super-Maxwell,
supersymmetry transformations of equation \eqref{eq:sp3_sym_susy}
up to a factor, since $Q_\al$ are also the super-Maxwell
supersymmetry generators. Combining the above two equations, we
learn
\begin{equation}
Q^a \psi_{g+1}^{LC} \sim \psi_{g+1}^{LC}(k_m ,
(0,{a^-}'({a^i}'),{a^i}'),({\chi^a}'({\chi^{\ad}}'),{\chi^{\ad}}')),
\end{equation}
where
\begin{eqnarray}
{a^i}'=\si^i_{a\bd}\chi^{\bd} && {\chi^{\ad}}' = -(2)^{\half} k^+
a^i \si^i_{a\bd}.
\end{eqnarray}
Therefore, using the map between $\psi_{g+1}^{LC}$ and $\psi_{BS}$
in section \ref{sec:gc7_BS_state_comparison}, we make the relation
\begin{eqnarray}\label{eq:gc6_S_a_Q_a}
Q^a \equiv i 2^{\frac{1}{4}} (P^{+})^\half S^a,
\end{eqnarray}
where $S^a$ describe the fermionic degrees of freedom for the
light-cone gauge, BS superparticle and where we have used equation
\eqref{eq:sp2_S_psi}.

The mapping between the $\Xh$'s of the two models is more
involved, so we simply provide the outline of a proof. To begin
with we relate our $\Kh^{mn}$, defined in equation
\eqref{eq:gc7_K_mn}, to gauge-invariant $(\Xh^m \Ph^n - \Xh^n
\Ph^m)$ of the semi-light-cone gauge, BS model. The $(\Xh^m \Ph^n
- \Xh^n \Ph^m)$ part of $\Kh^{mn}$ operates on $\psi_{g+1}^{LC}$
in an identical manner to how it operates on $\psi_{BS}$.
Unfortunately, however, all the terms in $\Kh^{mn}$ are necessary
in order that it be BRST closed. When $\Kh^{mn}$ operates on a
physical state $\psi_{g+1}(k_m,a_m,\chi^\al)$, it Lorentz rotates
the quantum numbers. We observe that $(\Xh^m \Ph^n - \Xh^n \Ph^m)$
Lorentz rotates $k_m$, while the remaining terms in $\Kh^{mn}$
Lorentz rotate $a_m$ and $\chi^\al$. We can build an operator
$\Sh^{mn}$ out of $\qh_\al$'s, which compensates for the rotation
of $a_m$ and $\chi^\al$, without rotating $k_m$. The $SO(8)$ part
of this term, for example, would be $\Sh^{ij} \sim \Sh^a
\si^{ij}_{ab}\Sh^b$, where $\Sh^a$ has been defined in terms of
$Q^a$ in equation \eqref{eq:gc6_S_a_Q_a}. Then
\begin{eqnarray}
\Rh^{mn}=\Kh^{mn} + \Sh^{mn}(\qh_\al) \label{eq:gc6_Rmn}
\end{eqnarray}
is exactly equivalent to $(\Xh^m \Ph^n - \Xh^n \Ph^m)$ of the
semi-light-cone, BS model.

It is fairly straightforward to relate $\Xh^i$ and $\Xh^-$ of the
light-cone gauge, BS model with $\Rh^{mn}$ of our model. We first
relate $\Xh^i$ and $\Xh^-$ to their gauge-invariant counterparts
\begin{eqnarray} \text{light-cone gauge BS}
&\longleftrightarrow& \text{gauge invariant with respect to $P^2=0$}\\
X^i &\longleftrightarrow& X^i - (P^+)^{-1}P^i(X^+ - \tau P^+)\\
X^- &\longleftrightarrow& X^- - (P^+)^{-1}P^-(X^+ - \tau P^+),
\end{eqnarray}
where the expressions on the left and right hand side are equal on
the light-cone gauge constraint surface. A convenient basis for
these operators is
\begin{equation}
\Ph^i, \Ph^+ , (\Xh^i \Ph^+ - \Xh^+ \Ph^i) , (\Xh^- \Ph^+ - \Xh^+
\Ph^-).
\end{equation}
Thus, any light-cone gauge, BS operator can be mapped to an
operator in our model, formed of the following basis elements
\begin{equation}\label{eq:gc6_basis_ops}
\Ph_m, \qh^a, \Rh^{i+} , \Rh^{-+}.
\end{equation}

To prove the reverse mapping for the $\Xh$'s is more difficult,
though given equation \eqref{eq:gc7_eff_constr_first_class}, it
seems reasonable to conjecture that every operator in our model
can be mapped to an operator in the light-cone gauge, BS
superparticle.

\section{Central charge cancellation for the open
superstring}\label{sec:gc7_quant_open_string}

In principle, the methods used in quantizing the superparticle
here can also be generalized to quantize the free superstring. It
is further confirmation of the fundamental nature of the Berkovits
BRST operator, combined with pure spinor ghosts, that the first,
excited massive, superspace vertex operator
\cite{Berkovits:2002qx} has been explicitly constructed, providing
for the first time the superspace form of the first massive
multiplet. Furthermore, the same principles can be used to
covariantly obtain the rest of the physical spectrum.

There are additional issues with the superstring which don't apply
to the superparticle. In particular, there is a quantum anomaly
which is the central charge in the Virasoro algebra. One expects
the central charge to disappear in $D=10$ as with the RNS
superstring.

The BRST charges are now
\begin{eqnarray}
Q= \oint{dz \la^\al(z) d_\al(z)}, && Q_{gc} =\oint{dz( C_m
\la\ga^m\la + \ldots)},
\end{eqnarray}
where we use the same notation as
Berkovits\cite{Berkovits:2000fe}, thus simply replacing world-line
parameter $\tau$ with complex, Euclidean world-sheet parameter
$z$.

The left-moving part of the superstring action is defined as
\begin{equation}
S=\int{d^2 z\,( \frac{1}{2}\pa X_m \pab{X}^m + \pab\theta^\al
p_\al + \pab\la^\al w_\al + \pab{C}_m B^m +  \pab U^\al V_\al +
...)},
\end{equation}
which is the open superstring version of equation
\eqref{eq:gc6_S_sp_full}. Hence, the energy momentum tensor is
given by
\begin{equation}
T_{zz}(z)=\frac{1}{2}\pa X_m \pa{X}^m + \pa\theta^\al p_\al  +
\pa\la^\al w_\al + \pa{C}_m B^m +   \pa U^\al V_\al + ... .
\end{equation}
The central charge contributions from $X$, $(p,\theta)$ and
$(w,\la)$ are +10 , -32, +32 respectively and from the ghost pairs
$(B,C)$, $(V,U)$, ... are -20, +32, ... . Each fermionic ghost
pair contributes -2 and each bosonic pair +2. From equation
\eqref{eq:cs3_graded_sum_m_i}, the graded sum of ghost constraint
ghost degrees of freedom, starting from $i=1$ instead of $i=0$, is
$-5$. Thus, the total contribution to the central charge by the
capital letter ghosts is $2\times (-5) = -10 $. The total central
charge is then
\begin{equation}
c= 10 -32 +32 -10 =0,
\end{equation}
as required, assuming that a termination for $Q_{gc}$ can be
found. If there are infinite ghosts for ghosts, $c$ will be an
infinite sum which must be regularized.


\section{Future research}\label{sec:gc8_future_research}

Either the ghosts for ghosts terms in $\Qh_{gc}$ have to be
completed, or some other method used before the ten dimensional
pure spinor and its conjugate momentum are covariantly quantized.
Only then will we have a complete, covariant BRST system for the
Berkovits superparticle.

Despite the above problem, we have seen in section
\ref{sec:gc7_pure_spinor_coho}, how it's still possible to
covariantly calculate matrix elements of arbitrary physical
operators, between physical states. Therefore it seems logical to
continue with the next step and build a model, in the same vein as
this paper, for the free Berkovits superstring, in the hope that
we can still perform useful calculations.

An outstanding problem is to derive tree-level superstring
scattering amplitudes. Although a plausible expression for
massless tree amplitudes has been conjectured, and tested
\cite{Berkovits:2000fe,Berkovits:2000ph}, it uses a special
integration measure, whose precise origin is unknown.
Understanding the origin of the tree-level amplitudes seems a
necessary step before we have a realistic chance of obtaining
one-loop amplitudes. It is hoped that understanding how to write
free superstring matrix elements, in a similar manner to the
superparticle in this article, will provide some insight towards
this goal. After all, in general we construct interacting string
amplitudes using the corresponding free string model.

\section*{Acknowledgements}

I would like to thank Nathan Berkovits, Marc Henneaux, and
especially Chris Hull and Dan Waldram for useful discussions. I'm
also grateful to Rosemary Reader for proof reading the text.
\appendix
\section{Conventions}
Roman letters in the middle of the alphabet $m,n,p$ etc.
correspond to space-time indices. Greek letters at the start of
the alphabet are used as spinor indices. The flat space-time metric
$\eta^{mn}$ has signature $-++\ldots +$. We also choose units such
that $c=1$ and $\hbar=1$.

Throughout, the graded Poisson bracket of functions $A$ and $B$ is
given by
\begin{equation}
[A,B].
\end{equation}
In the context of operators, which have hats,
\begin{equation}
[\Ah,\Bh],
\end{equation}
is the graded quantum (anti-)commutator of operators $\Ah$ and
$\Bh$. The brackets of generic, bosonic, conjugate pair $\Xh$ and
$\Ph$, and generic, fermionic, conjugate pair $\Ch$ and $\Bh$ are
given by
\begin{eqnarray}
[\Xh, \Ph]=i, && [\Ch,\Bh] = -i.
\end{eqnarray}
Also $\Xh$, $\Ph$ and $\Ch$ are hermitian, and $\Bh$ is
anti-hermitian. In the Schr\"odinger representation, $\Xh$ and
$\Ch$ are simply given by bosonic variable $X$ and fermionic
variable $C$ respectively. Similarly, their conjugate momenta
$\Ph$ and $\Bh$ are given by $-i \pa/\pa X$ and $-i \pa/\pa C$.

\section{BRST quantization in the Schr\"odinger
representation}\label{sec:BRST_Schro}

We state some useful results \cite{Henneaux:1992ig} regarding BRST
quantization in the Schr\"odinger representation.

The ghost number operator $\Gh$ is defined up to a constant, which
can be chosen such that it is anti-hermitian $\Gh=-\Gh^\dag$. We
then find that the inner product of two states $\psi_g$ and
$\psi_{g'}$ obeys
\begin{eqnarray}\label{eq:gcB_psi_g_psi_g_prime}
(\psi_g,\psi_{g'}) =0, && \text{for }g+g' \neq 0,
\end{eqnarray}
where $\Gh\psi_g=g\psi_g$ and $\Gh\psi_{g'}=g'\psi_g$. So for $g
\neq 0$, the state $\psi_g$ has zero norm and couples only to
states with ghost number $-g$. Also, there is a theorem that
opposite ghost number, state cohomologies are isomorphic
\begin{equation}
H^g_{st}(\Qh) \cong H^{-g}_{st}(\Qh),
\end{equation}
where $\Qh$ is the BRST operator.

In the Schr\"odinger representation, with no non-minimal sector
included, the physical state cohomology appears at ghost numbers
$\pm m/2$ for a standard gauge theory with $m$ irreducible, first
class constraints.

We therefore need to compute state cohomologies at both ghost
numbers in order to make matrix element calculations. They then
take the form $(\chi_{gc}, \Ah \psi_{-g})$, where $\psi_{-g}\in
H^{-g}(\Qh)$ and $\chi_{gc} \in H^{g}(\Qh)$ and where $\Ah$ is a
ghost number zero operator. For practical calculations, states in
each cohomology will be defined by a different, but equivalent set
of quantum numbers, which we therefore need to relate. We want an
explicit map between cohomology classes at the two ghost numbers.

We look for a basis $\{\psi^A_{g}\}$ for states in
$H^g_{st}(\Qh)$, and similarly a basis $\{\psi^B_{-g}\}$ for
states in $H^{-g}_{st}(\Qh)$, where $A$ and $B$ are indices, such
that
\begin{equation}
(\psi^B_{-g},\,\psi^A_g) = \de^{AB},
\end{equation}
and each cohomology class has just one representative which is a
linear combination of the basis elements. We then make the map
\begin{equation}\label{eq:gcB_coho_map}
\psi^A_{g}\longleftrightarrow \psi^A_{-g}.
\end{equation}

\section{D=10 Gamma matrices}\label{sec:gamma_matrices}
\subsection{Construction and basic properties}

The Dirac gamma matrices in ten dimensions are $32\times 32$
matrices $\Ga^m_{AB}$ obeying the Clifford algebra
\begin{equation}
\Ga^m_{AB} \Ga^n_{BC} + \Ga^n_{AB} \Ga^m_{BC} = 2
\eta^{mn}\de_{AC}.
\end{equation}
We choose the reducible, Majorana-Weyl representation, in which
$\Ga^m_{AB}$'s are real and consist of two symmetric $16\times 16$
matrices $\ga^m_{\al\be}$ and $\ga^{m\al\be}$ on the off-diagonals
\begin{equation}
\Ga^m =
\begin{pmatrix}
0 & \ga^{m\al\be} \\
\ga^m_{\al\be} & 0
\end{pmatrix}.
\end{equation}
In this notation, $\theta^\al$ is Weyl and $\theta_\al$ anti-Weyl,
thus a down spinor index can only be contracted with an up index
when building Lorentz covariant tensors. Since $\ga^m$ are real,
the Majorana condition simply says that
$\theta^\al={\theta^\al}^*$. We generally deal with only Weyl
spinors and hence use the $16\times 16$ $\ga^m$ notation.

The Clifford algebra in terms of $\ga^m$ reads
\begin{equation}\label{eq:B_ga_alg}
\ga^m_{\al\be} \ga^{n\be\de} + \ga^n_{\al\be} \ga^{m\be\de} = 2
\eta^{mn} \de^{\de}_\al .
\end{equation}
The $16\times 16$ gamma matrices can be built from the $SO(8)$
gamma matrices which themselves are direct products of Pauli
matrices \cite{Green:1987sp}. The antisymmetric, real $SO(8)$
Pauli matrices $\{\si^i_{a\bd}\,i=1,\ldots,8\}$, which obey the
Clifford algebra
\begin{equation}
\si^i_{a \ad}\si^j_{\ad b} + \si^j_{a \ad}\si^i_{\ad b} =
2\de^{ij} \de_{a b},
\end{equation}
can be used to construct $\ga^m_{\al\be}$. Specifically
\begin{equation}
\ga^i_{\al\be}=
\begin{pmatrix}
0 & \si^i_{a\ad}\\
\si^i_{\bd b} & 0
\end{pmatrix},
\end{equation}
where $i=1,\ldots,8$. We define $\ga^{m\al\be}$ by exactly the
same expression. We see that a Weyl spinor splits as $\theta^\al=
(\theta^a , \theta^{\ad})$. A ninth matrix which anticommutes with
these eight is given by $\ga^9_{\al\be} =
\ga^1_{\al\be_1}\ga^{2{\be_1}{\be_2}}\ldots \ga^8_{\be_7\be}$,
which given the $SO(8)$ matrices we can calculate below. The
values of $\ga^0_{\al\be}$ and $\ga^{0\al\be}$ are similarly
defined in order to be consistent with their algebra
\eqref{eq:B_ga_alg}
\begin{eqnarray}
\ga^9_{\al\be} =\ga^{9\al\be}=
\begin{pmatrix}
1 & 0\\
0 & -1
\end{pmatrix}\\
\ga^0_{\al\be} =
\begin{pmatrix}
-1 & 0\\
0 & -1
\end{pmatrix} &&
\ga^{0\al\be} =
\begin{pmatrix}
1 & 0\\
0 & 1
\end{pmatrix}.
\end{eqnarray}

A generic antisymmetric product of $r$ $\ga^m$'s is notated as
\begin{equation}
\ga^{m_1 m_2 ... m_r} = \ga^{[m_1} \ga^{m_2} ... \ga^{m_r]},
\end{equation}
where a factor of $1/r!$ is implicit, remembering that a
$\ga^{m_1}_{\al\be}$ must contract with a $\ga^{{m_2}\be\de}$
etc.. This larger set of gamma matrices, defined by the full set
of antisymmetric combinations, form a basis for bispinors. There is
a duality
\begin{equation}
\ga^{m_1 m_2 ... m_r} = \frac{1}{(10-r)!}\eps^{m_1 ... m_r m_{r+1}
... m_{10}}\ga_{m_{r+1}} \ldots \ga_{{m_{10}}},
\end{equation}
in particular, $\ga^{mnpqr}$ is self-dual, so only half of the
$\ga^{mnpqr}$'s are independent. A generic bispinor with either 2
lower or 2 upper indices is a linear combination of $\ga^m$,
$\ga^{mnp}$ and $\ga^{mnpqr}$. For example
\begin{equation}\label{eq:C_bispinor}
f_{\al\be} = f_m \ga^m_{\al\be} + f_{mnp}\ga^{mnp}_{\al\be} +
f_{mnpqr}\ga^{mnpqr}_{\al\be},
\end{equation}
where $f_m$, $f_{mnp}$ and $f_{mnpqr}$ are calculated in terms of
$f_{\al\be}$ by using the orthogonal properties of the gamma
matrices. For example $f_m = \ga_m^{\al\be} f_{\al\be}/16$.

From the definition, $\ga^m$ and $\ga^{mnpqr}$ are symmetric,
while $\ga^{mnp}$ is antisymmetric. Similarly, ${\de^{\al}}_\be$,
${\ga^{mn\al}}_\be$ and ${\ga^{mnpq\al}}_\be$ form a basis for
bispinors with one lower and one upper index. The matrices
${\de^{\al}}_\be$, ${\ga^{mnpq\al}}_\be$ are symmetric, while
${\ga^{mn\al}}_\be$ is antisymmetric.

\subsection{Gamma matrix identities}

In principle, all identities can be calculated from the Fierz
identity and the Clifford algebra identity given below
\begin{eqnarray}\label{eq:B_Fierz_Clifford}
\ga_{m\al(\be}\ga^m_{\ga\de)} = 0, &&
\ga^{(m}_{\al\be}{\ga^{n)}}^{\be\de} = \eta^{mn}\de_\al^\de,
\end{eqnarray}
though in practice, this is far too time consuming for all but the
most simple identities. A common requirement is to calculate a
product of $\ga$'s in terms of a sum of $\ga$'s. For this purpose,
a slicker method is to use Young tableaux, which are useful for
determining direct products of tensors in terms of sums of tensors
with definite (anti)-symmetry properties in their indices. For
example
\begin{equation}
\ga^m \ga^{npqr}= k_1 \ga^{mnpqr} + k_2 \eta^{m[n} \ga^{pqr]},
\end{equation}
where $k_1$ and $k_2$ are constants, and we have used that
$\ga^{(m} \ga^{n)}=\eta^{mn}$. We then calculate $k_1$ and $k_2$
by substituting particular values of $m,n,p,q,r$ into the above
equation. In this case $k_1=1$ and $k_2=4$. Some more useful
identities are
\begin{eqnarray}
\ga^{m\al\be}\ga^n_{\al\be} = 16 \eta^{mn}, &&
\ga^m_{\al\be}\ga_m^{\be\de} = 10\de_\al^\de,
\end{eqnarray}
\begin{equation}
\ga^m \ga^{np} = \ga^{mnp} + 2\eta^{m[n} \ga^{p]}.
\end{equation}

\section{Description of pure spinors using U(5) co-ordinates}\label{sec:sp7_U5}

By first Wick-rotating from $SO(9,1)$ to $SO(10)$ and using $U(5)$
co-ordinates, we can parameterize the pure spinor constraint
surface non-degenerately in a certain co-ordinate patch.

Using Berkovits' notation,
\begin{eqnarray}
X^a = (X^1 + iX^2), \ldots , ( X^9 + iX^{10}), && a=1,\ldots,5,\\
X_a = {X^a}^\dag = (X^1 - iX^2), \ldots, (X^9 - i X^{10}),
\end{eqnarray}
where $X^{10} = -i X^0$. So $X^a$ and $X_a$ transform in
the $\textbf{5}$ and $\bar{\textbf{5}}$ representation of the
$U(5)$ group. Thus, we define the $U(5)$ gamma matrices $\ga^a$ and
$\ga_a$ in the same manner, except we include a normalization
factor so that $\ga^a = (\ga^1+i\ga^2)/\sqrt{2}$. The gamma matrix
algebra is
\begin{eqnarray}
\{ \ga^a , \ga^b\} = \{ \ga_a , \ga_b\} = 0, && \{ \ga^a, \ga_b \}
= 2\de^a_b,
\end{eqnarray}
so we can treat $\ga^a$ as raising and $\ga_a$ as lowering
operators in order to create a generic spinor. For example, the
ground state spinor $u_+^\al$ is defined by $\ga_a u_+=0$ for
$a=1,\ldots,5$. By acting with up to 5 $\ga^a$'s on the ground
state $u_+$, we obtain the full set of spinors. We see that acting
with an odd number of $\ga^a$'s on $u_+^\al$ changes the
chirality, since an up $\al$ index can only contract with a down
$\al$ index. For example $u^\al_+$ and $\ga^a_{\al\be} u^\be_+$
have opposite chirality. A basis for the spinor $\la^\al$ is given
by $u_+$, $(u^{ab})^{\al}= \ga^a \ga^b u_+$ and $u_a^\al =
\eps_{abcde}\ga^b\ga^c\ga^d\ga^e u_+$. Thus
\begin{equation}\label{eq:sp7_U5_lambda}
\la^\al = \la^+ u_+^\al + \la_{ab}(u^{ab})^\al + \la^a u_a^\al,
\end{equation}
where $(\la^+, \la_{ab}, \la^a)$ transform in the
$(\textbf{1},\bar{\textbf{10}}, \textbf{5})$ representation of
$U(5)$. It's a similar story for a spinor of opposite chirality,
like $w_\al$, except that the basis comes from applying 1, 3 and 5
$\ga^a$'s respectively to $u_+^\al$. The spinor $w_\al$ splits
into $(w_+, w^{ab}, w_a)$, a $(\bar{\textbf{1}}, \textbf{10},
\bar{\textbf{5}})$ representation of $U(5)$. In calculating
$\la^\al w_\al$ for example in terms of $U(5)$ co-ordinates, we
use the result that $u_+ \ga^1\ga^2\ga^3\ga^4\ga^5u_+=1$ and that
$u_+ \ga^a \ga^b \ga^c u_+$ and $u_+ \ga^a u_+$ vanish. Thus, the
pure spinor constraints become
\begin{eqnarray}\label{eq:sp7_lgl_U(5)1}
\la \ga^a \la = \la^+ \la^a + \frac{1}{8}\eps^{abcde}\la_{bc}\la_{de}=0,\\
\la \ga_a \la = \la^b \la_{ab}=0 \label{eq:sp7_lgl_U(5)2}.
\end{eqnarray}
In the region defined by $\la^+ \neq 0$, we use equation
\eqref{eq:sp7_lgl_U(5)1} to write $\la^a$ in terms of $\la_+$ and
$\la_{ab}$
\begin{equation}\label{eq:sp7_la}
\la^a = \frac{-1}{8}(\la^+)^{-1}\eps^{abcde}\la_{bc}\la_{de}=0.
\end{equation}
We then find that the expression for $\la^a$ in equation
\eqref{eq:sp7_la} automatically satisfies the second condition
\eqref{eq:sp7_lgl_U(5)2}, since, using the Young tableaux
expression for tensors of specific symmetry properties, $\la_{a[b}
\la_{cd}\la_{ef]}=k_1 \la_{[ab} \la_{cd}\la_{ef]} + k_2
\la_{(a[b)} \la_{cd}\la_{ef]} =0$, where $k_1$ and $k_2$ are
constants, and $\la_{(ab)}=0$.

Since the ghosts $\la^\al$ are constrained, leaving 11 free
complex parameters, i.e. $\la^\al=\la^\al(\la^+, \la_{ab})$, we
expect some suitable constraints to be placed on $w_\al$, such
that it also has 11 complex degrees of freedom. Firstly, note that
if we treat $\la\ga^a\la=0$ as first class constraints, recalling
that $\la\ga_a\la=0$ are redundant for $\la^+ \neq 0$, the
variation of $w_a$ under a gauge transformation is given by
\begin{equation}
\de w_a = -\eps_a \la^+,
\end{equation}
where $\eps_a(\tau)$ is a local, bosonic parameter. Therefore, a
good canonical gauge, which is both accessible and completely
fixes the gauge symmetry, is given by
\begin{equation}
w_a=0.
\end{equation}
The constraints $\la\ga^a\la=0$ and $w_a=0$ describe a second class
constraint surface in the region $\la^+\neq 0$. We parameterize the
constraint surface using the co-ordinates $\la^+$, $\la_{ab}$,
$w_+$, $w_{ab}$. The induced, Poisson bracket between the
co-ordinates of the constraint surface needs to be calculated. The
simplest way to calculate the bracket is to write the ghost action
for the superparticle
\begin{equation}
S_g = \int{d\tau (\dot{\la}^a w_a + \dot{\la}^+ w_+ +
\half\dot{\la}_{ab}w^{ab} - \La_a \la\ga^a\la - \La^a
\la\ga_a\la)},\end{equation}
then parameterize the second class
constraint surface with $\la^+, \la_{ab},w_+,w_{ab}$ so that
\begin{equation} S_g = \int{d\tau (\dot{\la}^+ w_+ +
\half\dot{\la}_{ab}w^{ab} )}. \end{equation}
It is clear to see that $(\la^+ , w_+)$ and $(\la_{ab},w^{ab})$
are two conjugate pairs. The bracket between $\la^\al$ and
$w_\be$ is given by
\begin{equation}\label{eq:sp7_U5_D_bracket}
[\la^\al(\la^+,\la_{ab}) ,w_\be(w_+,w^{ab})]_* = \de^\al_\be -
u_a^\al v^a_{\be},
\end{equation}
where $[\cdot,\cdot]_*$ is the induced Poisson bracket on the
constraint surface, and where $u^\al_a$ and $v_\be^b$ are defined
as basis spinors for $\la^a$ and $w_b$ respectively as in equation
\eqref{eq:sp7_U5_lambda}.


Instead of parameterizing the constraint surface, we could have
alternatively defined a Dirac bracket.

\section{D=10, N=1, Brink-Schwarz superparticle
and super-Maxwell theory}\label{sec:gc2_review}
\subsection{D=10, N=1, Brink-Schwarz superparticle in the
semi-light-cone gauge} \label{sec:sp2_superp}

For brevity, we directly specify the Brink-Schwarz superparticle
in the semi-light-cone gauge, with no derivation. In this gauge,
the fermionic, $\kappa$ symmetry is gauge-fixed, but not the
world-line reparameterization symmetry.

The system is defined by fundamental, bosonic operators $\Xh^m$,
$\Ph_m$, and fermionic operators $\Sh^a$, $a=1,\ldots,8$,
whose quantum commutator algebra is
\begin{eqnarray}\label{eq:sp2_S_Cliff_alg} \Xh^m \Ph_n - \Ph_n\Xh^m
= i\de^m_n, && \Sh^a \Sh^b + \Sh^b \Sh^a = 2\de_{ab},
\end{eqnarray}
where all other brackets are zero. There is also the first class
constraint $\Ph_m\Ph^m=0$, which generates the world-line
reparameterization symmetry and the Hamiltonian is $\Ph_m\Ph^m/2$.

So $\Sh^a$ form a Clifford algebra and a representation can be
built from the SO(8) Pauli matrices $\si^i_{a\bd}$, described in
appendix \ref{sec:gamma_matrices}. A generic wavefunction
$\psi_{BS}(X)$ in the representation space is
\begin{eqnarray}
\psi_{BS}(X) = e^{ik.X}(\eps^i|i> + \eps^{\ad}|\ad>), && k^m k_m =
0,
\end{eqnarray}
where $\eps^i$, $\eps^{\bd}$ are bosonic constants, $\eps^i$ being
a spin 1 $SO(8)$ vector, and $\eps^{\ad}$ a spin $1/2$
anti-chiral, SO(8) spinor, and where $i,\ad =1,\ldots,8$. Also
states $|i>$ and $|\ad>$ are normalized as $<i|j>=\de_{ij}$, $<\ad
|\bd>=\de_{\ad\bd}$, and $\Sh^a$ acts on $\psi_{BS}$ as follows
\begin{equation}\label{eq:sp2_S_psi}
\Sh^a \psi_{BS}(X) =e^{ik.X} (\si^i_{a\bd}\eps^{\bd}|i> +
\si^i_{a\bd}\eps^i|\bd>).
\end{equation}
The inner product between two physical states $\psi_{BS1}$ and
$\psi_{BS2}$ is given by
\begin{eqnarray}\label{eq:sp2_BS_sc_prod}
(\psi_{BS1} , \psi_{BS2}) \propto ((\eps_1^i)^* \eps_2^i +
(\eps_1^{\bd})^*\eps_2^{\bd}) \de^{(10)}(k_1 - k_2).
\end{eqnarray}
We can show that $S^a= i2(2)^{\frac{1}{4}}(P^+)^\half\theta^a$,
where $\theta^{\al}=(\theta^a, \theta^{\ad})$ is the usual
fermionic, superspace variable and $P^\pm= (P^0 \pm
P^9)/\sqrt{2}$.

The wavefunction $\psi_{BS}$ corresponds, up to normalization
constants, to the light-cone gauge, classical field multiplet of
$D=10$, $N=1$ super Maxwell theory
\begin{equation}\label{eq:sp2_psi_BS_LC}
\psi_{BS} = a^i|i>  -i2^{\frac{1}{4}}(P^+)^{-\half}\chi^{\bd}|\bd>,
\end{equation}
where $a^i$ and $\chi^{\bd}$ behave like the light-cone gauge
photon and photino fields of super-Maxwell. The normalization
factor is included \cite{Green:1987sp}, so that the super-Maxwell
and the BS superparticle supersymmetry transformation exactly
coincides.

\subsection{D=10, N=1 Super-Maxwell}
\subsubsection{The action, symmetries and
light-cone gauge fields}\label{sec:sp3_S_S_Y_M}

The SO(9,1) covariant action is the usual super Yang-Mills one,
with $U(1)$ gauge group
\begin{equation}\label{eq:sp3_S_Y_M}
S = \int{d^{10}X \, (\frac{i}{2}\chi\ga^m \pa_m\chi -\frac{1}{4}
f^2}),
\end{equation}
where $\chi^\al$ is a Majorana-Weyl spinor, and where
$f_{mn}=-ig[\pa_m a_n - \pa_n a_m]$ is the field strength for
the $U(1)$ gauge field $a_m$.
The infinitesimal gauge symmetry is
\begin{eqnarray}
\de \chi^\al = 0 && \de a_m = g \pa_m \phi(X),
\end{eqnarray}
and the action possesses the following supersymmetry,
\begin{eqnarray}\label{eq:sp3_sym_susy}
\de \chi^\al = - f_{mn}(\ga^{mn} \eps)^\al && \de a_m = i (\eps
\ga_m \chi),
\end{eqnarray}
for infinitesimal, fermionic, Majorana-Weyl constant $\eps$.

To describe the physical modes, we choose the light-cone gauge
$\pa_m a^m =0$, $a^+ = 0$, where we assume again that $p^+\neq 0$.
In momentum space
\begin{eqnarray}
a^- = \frac{1}{p^+} p^i a^i, && \pa^2 a^i =0,
\end{eqnarray}
thus the 8 massless, transverse modes $a^i$, describe the bosonic,
physical sector. Also, $\chi^\al$ obeys the Dirac equation $\ga^m
p_m \chi = 0$, which can be written as
\begin{eqnarray}\label{eq:sp3_chi_a_from_chi_adot}
\chi^a = -\frac{1}{(2)^{\half}p^+}p^i\si^i_{a \bd} \chi^{\bd}, &&
\pa^2 \chi^{\ad} = 0,
\end{eqnarray}
leaving the 8 massless modes $\chi^{\ad}$, which describe the
fermionic physical sector.

\subsubsection{The superspace formulation}\label{sec:sp3_S_Y_M_superspace}

For details of $D=10$, $N=1$ superspace, super Yang-Mills see
\cite{Witten:1986nt}. We only specify results relevant to this
work.

The constraint equation is $F_{\al\be}=0$, where $F_{\al\be}$ is
the spin one superfield strength
\begin{equation}\label{eq:sp3_Y_M_constraints}
F_{\al\be}= D_\al A_\be + D_\be A_\al + 2i \ga^m_{\al\be} A_m,
\end{equation}
and where $D_\al$ is the covariant superspace derivative of
equation \eqref{eq:gc2_cov_derivative}, and $(A_\al,A_m)$ the
superspace, gauge connection. Since $F_{\al\be}$ is a symmetric
bispinor, it can be determined in terms of symmetric gamma
matrices of the same chirality $F_{\al\be} = F_m\ga^m_{\al\be} +
F_{mnpqr}\ga^{mnpqr}_{\al\be}$, as in appendix
\ref{sec:gamma_matrices}. Thus, the constraint equations split
into two pieces, the first of which, $F_m=0$, simply determines
$A_m$ in terms of $A_\al$, and the second of which, $F_{mnpqr}=0$,
implies the equations of motion for $A_\al$
\begin{equation}\label{eq:sp3_A_al_constraint_eqns}
\ga^{\al\be}_{mnpqr}D_{\al} A_{\be} =0,
\end{equation}
which have the effect of placing the theory on-shell. The
equations of motion for the super field-strengths can be deduced
from the Bianchi identities and the constraint equation
\cite{Witten:1986nt}. They are
\begin{eqnarray}
\ga^m_{\al\be} \pa_m W^\be = 0, && \pa^m F_{mn} = 0,
\end{eqnarray}
where $W^\al$ is the photino superfield-strength, given by
$W^\al=(\ga^m)^{\al\be}F_{m\be}/10$, and where $F_{mn}=\pa_m A_n
-\pa_n A_m$ and $F_{\al m}= D_\al A_m - \pa_m A_\al$. The photino
equation is equivalent to $\pa^m F_{m\al}=0$.

In a $\theta^\al$ expansion of the field strengths, the zero
components are $F^{mn}|_{\theta=0}=f^{mn}$ and
$W^\al|_{\theta=0}=\chi^\al$. We can choose a gauge, using $\de
A_\al = D_\al \phi$, such that
\begin{equation}\label{eq:sp7_A_al_cmpts}
A_\al = -i a_m(X) \ga^m_{\al\be} \theta^\be - \chi^{\ga}(X)
\ga^m_{\al\be}\ga_{m\ga\de} \theta^\be \theta^\de + \ldots,
\end{equation}
where $a_m$ and $\chi^\al$ obey the super-Maxwell equations of
motion in the Lorentz gauge, $\pa^2 a_m = \pa^m a_m =0$ and
$\ga^m_{\al\be}\pa_m \chi^{\be}=0$. The $\ldots$ denote terms at
higher order in $\theta$ which depend on space-time derivatives of
$a_m$ and $\chi^\al$.

\bibliographystyle{utphys}
\bibliography{xbib}
\end{document}